\documentclass[twocolumn]{aastex63}
\usepackage{amsmath}
\usepackage{comment}

\usepackage{enumitem}

\newcommand{\be}{\begin{equation}}
\newcommand{\ee}{\end{equation}}

\shorttitle{Long term evolution of massive-star, post-CE circumbinary disks}
\shortauthors{Semih Tuna, B.~D.~Metzger}
\graphicspath{{./}}

\begin{document}

\title{Long-Term Evolution of Massive-Star Post-Common Envelope Circumbinary Disks and the Environments of Fast Luminous Transients}

\author[0000-0002-8304-1988]{Semih Tuna}
\affil{Department of Physics and Columbia Astrophysics Laboratory, Columbia University, New York, NY 10027, USA}

\author[0000-0002-4670-7509]{Brian D.~Metzger}
\affil{Department of Physics and Columbia Astrophysics Laboratory, Columbia University, New York, NY 10027, USA}
\affil{Center for Computational Astrophysics, Flatiron Institute, 162 5th Ave, New York, NY 10010, USA}

\correspondingauthor{Semih Tuna}
\email{semih.tuna@columbia.edu}

\begin{abstract}
If the envelope of a massive star is not entirely removed during common envelope (CE) interaction with an orbiting compact (e.g., black hole [BH] or neutron star [NS]) companion, the residual bound material eventually cools, forming a centrifugally-supported disk around the binary containing the stripped He core.  We present a time-dependent height-integrated model for the long-term evolution of post-CE circumbinary disks (CBD), accounting for mass and angular momentum exchange with the binary and irradiation heating by the He core and photoevaporation wind mass-loss.  A large fraction of the CBD's mass is accreted prior to its outwards viscous spreading and wind-dispersal on a timescale $\sim 10^{4}-10^{5}$ yr, driving significant changes in the binary separation, even for disks containing $\sim 10\%$ of the original envelope mass.  Insofar that the CBD lifetime is comparable to the thermal (and, potentially, nuclear) timescale of the He core, over which a second mass-transfer episode onto the companion can occur, the presence of the CBD could impact the stability of this key phase.  Disruption of the He core by the BH/NS would result in a jetted energetic explosion into the dense gaseous CBD ($\lesssim 10^{15}$ cm) and its wind ($\gtrsim 10^{16}$ cm), consistent with the environments of luminous fast blue optical transients like AT2018cow.  Evolved He cores which undergo core-collapse still embedded in their CBD could generate Type Ibn/Icn supernovae.  Thousands of dusty wind-shrouded massive-star CBD may be detectable as extragalactic luminous infrared sources with the {\it Roman Space Telescope;} synchrotron radio nebulae powered by the CBD-fed BH/NS may accompany these systems.

\end{abstract}


\section{Introduction}
\label{sec:introduction}

The interaction of massive star binaries guides the evolutionary channels which give rise to many astrophysical stellar populations, including short-period stellar binaries (e.g., \citealt{Meyer&Meyer-Hofmeister79,Kalogera&Webbink98,Han+95,Belczynski+16}), the progenitors of thermonuclear \citep{Webbink84} and stripped-envelope supernovae (e.g., \citealt{Iben&Tutukov84,Podsiadlowski+92,Eldridge+08,Yoon+10,Zapartas+19,Sravan+19}) as well as LIGO (e.g., \citealt{Abbott+17,Vigna-Gomez+18}) and LISA gravitational wave sources (e.g., \citealt{Ablimit+16,Chen+20,Renzo+21}).  These systems host evolved stars found at orbital separations much smaller than their sizes earlier in their evolution, thus raising the question of what brought them so close to each other.  A promising formation channel for such short-period binaries is a brief epoch of binary evolution known as the ``common envelope" (CE) phase (\citealt{Paczynski76}, see \citealt{Ivanova+13, Ivanova+20} for a review). The CE occurs after a donor star fills its Roche Lobe, initiating unstable mass transfer onto its companion . The expelled mass cannot be accommodated by the accretor due to the runaway nature of the process, instead forming a gaseous envelope that simultaneously engulfs the core of the donor and the accretor.  Eventually the accretor loses corotation with the envelope and spirals in towards the core, imparting orbital energy and angular momentum to the envelope (e.g., \citealt{Morris&Podsiadlowski06}).  Depending on the initially available energy, and the efficiency with which it is used to unbind the envelope, the latter may be completely ejected, leaving the companion in a tight orbit around the stripped core of the accretor.

CE evolution is one of the least understood phases of stellar evolution due to the complexity and variety of physical processes operating across a wide range of temporal and spatial scales. The rapid (dynamical timescale) spiral-in phase ($\lesssim \rm decades$) is typically preceded by a potentially long \citep{Soker15, Roepke&DeMarco22} mass-transfer phase that sets the initial conditions for the CE \citep{Ivanova+13}, and has been studied due to its role in affecting the appearance of red luminous transients \citep{Pejcha+16b,Pejcha+16a,Pejcha+17,Metzger&Pejcha17, Blagorodnova+21}. Loss of mass and angular momentum during this phase can lead to varying initial conditions for the inspiral, in turn imparting variety to the outcomes \citep{MacLeod&Loeb20}.  Hydrodynamic simulations of CE interaction show that the binary later enters another long, self-regulated spiral-in phase \citep{IvanovaTaam04}, though the existence of this phase has been questioned \citep{Lau+22a, Lau+22b}. Lack of spatial symmetries and the necessity to follow spatial and temporal scales spanning orders of magnitude \citep{Ivanova+13} render the problem analytically intractable, thus limiting most numerical studies to the rapid dynamical phase (see \citealt{Roepke&DeMarco22} for a review).  A fully satisfactory model which relates the binary's post-CE state to its pre-CE initial conditions is not available to this day.

A better understanding of the final separations of the binary following the CE phase is crucial, for instance, to making accurate predictions for the rates of LIGO compact object mergers (e.g., \citealt{Dominik+12,Broekgaarden+19,Broekgaarden+21,Marchant+21,Klencki+21}). Analytical approaches to this problem are limited by simple parameterizations of the energy/angular momentum budget (so called $\alpha$ \citep{Webbink84} and $\gamma$ \citep{Nelemans+00} formalisms, respectively), in which a fixed fraction of the pre-CE orbital energy/angular momentum is assumed to be used to unbind the donor envelope; the final orbital separation is determined by global conservation of these quantities, bypassing the complicated dynamics of the CE phase. Although the $\alpha$ formalism is widely used in population synthesis studies applied to observations (e.g., \citealt{Hurley+02}), its reliability is questionable because of the unknown dependence of $\alpha$ on the binary and stellar properties and other observations which challenge its predictions \citep{DeMarco+11}.

The post-CE orbital separation is closely related to the state of the envelope after the spiral-in phase, another open problem in CE evolution. One conclusion of modern three-dimensional hydrodynamic CE simulations  \citep{Ricker&Taam08,Ricker&Taam12,Passy+12,Nandez+15,Ohlmann+16,Ivanova&Nandez16,MacLeod+18,Chamandy+18,Chamandy+20,Reichardt+19,Prust&Chang19,Reichardt+20,Sand+20,Glanz&Perets21,Moreno+22,Lau+22a,Lau+22b} is that the envelope may fail to be removed, motivating additional physical processes not captured by present simulations (e.g., \citealt{Nandez+15,Sabach+17,Glanz&Perets18,Wilson&Nordhaus19}). If and when such ``failures'' occur, one possible outcome is a dynamical-timescale merger between the accretor and the He core of the original donor.  While not leaving a surviving compact binary, such mergers may still give rise to luminous accretion-powered explosions inside the dense remaining hydrogen envelope (e.g., \citealt{Fryer&Woosley98,Chevalier12,Soker&Gilkis18,Soker+19,Schroder+20,Grichener&Soker21}).\footnote{Or in case of mergers with non-degenerate stars, producing a star which will undergo core-collapse supernova significantly later than expected for single-star evolution (e.g., \citealt{Zapartas+17})}. Alternatively, those portions of the envelope which are not removed may end up in a centrifugally supported, rotating ``disk'' surrounding the He core/accretor binary (e.g., \citealt{Sandquist+98,Kashi&Soker11,Lau+22a,Lau+22b}): a post-CE circumbinary disk (CBD). The system's ultimate fate then depends on long-term mass and angular momentum transfer between the CBD and the binary.  

 At late stages in the evolution of interacting binaries comprised of lower-mass stars (so-called ``post-AGB'' or ``post-red giant'' systems), an abundance of data supports the presence of long-lived dusty Keplerian disks (e.g., \citealt{Bujarrabal+13,Bujarrabal+18}; see \citealt{vanWinckel18} for a review).  These CBDs may be relics of a CE or CE-like phase, similar to those hypothesized to power jets observed during in proto-planetary nebulae (e.g., \citealt{Soker02,Frank&Blackman04,Ondratschek+22}).  
 
 Motivated by early findings of a bound post-CE envelope \citep{Sandquist+98, DeMarco+11}, \citet{Kashi&Soker11} found that such post-CE CBD could drive significant orbital shrinkage and excite orbital eccentricity.  For a $0.6\,M_\odot$ white dwarf orbiting a post-AGB core of mass $0.77 M_\odot$ on a circular orbit at $a \approx 1.4\,R_\odot$ following the CE phase, they showed that resonant interactions \citep{Artymowicz+91} with a CBD of mass $M  \gtrsim 0.16\,M_\odot$ can lead to a merger in less than a year, whereas a less massive disk would only drive the binary to merge in $\gtrsim 10^6\,\rm yr$ or longer.  Resonant interactions have been similarly invoked to explain the significant observed eccentricities $e \sim 0.1-0.5$ of short-period post-AGB binaries \citep{Dermine+13}. \citet{Izzard&Jermyn18,Izzard&Jermyn23} developed a rapid population synthesis algorithm, modelling the CBD - post-AGB binary mass and angular momentum transfer, assuming a steady, thin disk model coupled to binary evolution at late stages of the CE phase.  Similar population synthesis models that account for mass and angular momentum transfer between the binary and a ``circumbinary ring" have been employed in the context of Be X-ray binaries and gravitational wave sources \citep{Portegies95, Mennekens&Vanbeveren14, Vinciguerra+20, vanSon+22}. \citet{Gagnier&Pejcha22} performed 3D hydrodynamic simulations of the early evolution (hundreds of orbits) of post-CE envelope surrounding the post-CE binary, demonstrating its resemblance to a CBD whose extrapolated influence was to induce significant orbital shrinkage in $\sim 10^3 - 10^5$ orbital periods.
 
 The existence of rotationally supported post-CE envelopes may be not limited to relatively low-mass $M \gtrsim M_\odot$ post-AGB systems, but are also seen in hydrodynamic simulations of more massive stars, e.g., red super giants with $M \gtrsim 10\,M_\odot$ \citep{Lau+22a, Lau+22b}.  Hence, many of the considerations applied to post-AGB disk-binary interaction can in principle be extended to massive CE events.  Depending on the interplay between factors such as the disk lifetime, the timescales for internal evolution of the stripped He core, the post-CE binary separation, and the strength of disk-binary interactions, the disk could drive significant orbital migration and provide a dense, aspherical distribution of mass in the environment of a merger-driven or core-collapse supernova explosion.  A post-CE CBD phase, therefore, has important consequences for the electromagnetic signatures of these events. Indeed, such massive post-CE systems with CBD that survive for thousands of years (or longer) and extend to radial scales $\sim 10^{15}\,\rm cm$ were proposed by \citet{Metzger22} as progenitors of luminous fast blue optical transients (LFBOTs) \citep{Drout+14,Arcavi+16,Prentice+18,Margutti+19,Ho+21b}, in particular, the well-studied event AT2018cow \citep{Prentice+18,Perley+19,Margutti+19,Ho+19}. This motivates a detailed study of the long timescale ($\sim 10^3 - 10^6\,\rm yr$) evolution of post-CE CBD and their large-scale wind-fed environments.
 
 In this work, we develop a parametrized, time-dependent, height-integrated model for the long-term viscous evolution of massive-star, post-CE CBD coupled to their binary's orbital evolution through mass and angular momentum transfer. The system we have in mind are in many ways similar to a variety of other astrophysical CBD, including post-AGB disks, massive protostellar disks \citep{Artymowicz+91,Bate&Bonnell97,Armitage2011,Matsumoto+19}, CVs (e.g., \citealt{Taam&Spruit01,Dubus+02}), and supermassive black hole binaries embedded in AGN disks \citep{Armitage+Natarajan02,MacFadyen&Milosavljevic08,Noble+12,LopezArmengo+21}; and much of the knowledge from these fields can be imported to this novel context \citep{Gagnier&Pejcha22}.  
 
We use our model to address several key questions: 
 \begin{enumerate}[label=(\roman*)]
 \item What impact does a CBD of given initial properties have on the orbital evolution of the He core/accretor binary? Under the time constraints imposed by the core evolution, how much mass accretion and orbital migration is possible?  What fraction of the disk mass is instead lost to winds?
 \item If the CBD can survive until a second phase of mass-transfer from the donor star (this time, its stripped core) onto its companion (perhaps due to loss of orbital angular momentum to the disk), how will the disk-binary coupling affect the stability of mass transfer in this phase?  
 \item What are the implications of a post-CE CBD for the merger-induced explosion or the core-collapse supernova explosion? In particular, is the \emph{delayed merger scenario} \citep{Metzger22} proposed for LFBOTs consistent with the observations probing the environment of, e.g. AT2018cow or Type Ibn/Icn supernovae?  
 \item What are other directly observable features of long-lived CBD in massive-star post-CE systems?  Can such rare objects be observed in extragalactic surveys through their dust-reprocessed or radio jetted emissions?
 \end{enumerate}
This paper is organized as follows.  In Section \ref{sec:setup} we overview the problem setup and motivate through analytic the key properties of the CBD formed from common envelope systems. In Section \ref{sec:model} we describe the equations solved and the detailed setup of the model. In Section \ref{sec:results} we present our results, several consequences of which are addressed in Section \ref{sec:discussion}.  We summarize our conclusions in Section \ref{sec:conclusions}.

\section{Motivation and Physical Setup}
\label{sec:setup}

\subsection{From Common Envelope to Circumbinary Disk}

We consider an initial binary system consisting of an evolved star of mass $M_{\star} \gtrsim 10M_{\odot}$ and radius $R_{\star}$ which has initiated Roche lobe overflow (RLOF) onto a compact companion of mass $M_{\bullet}$.  Since we are in part interested in the production of relativistic transient sources, we focus on cases where the companion is a neutron star (NS) or black hole (BH) of characteristic mass $M_{\bullet} \sim 1.4-10M_{\odot}$, formed from an earlier stellar core-collapse event (likely also preceded by binary interaction; e.g., \citealt{Grichener23}).  However, some of our results are not sensitive to the exact nature of the companion, which could instead be an unevolved (e.g., main-sequence) star (e.g., \citealt{Gotberg+18}).  

The orbital semi-major axis at RLOF is given by \citep{Eggleton83}
\be
a_{\rm RLOF} \approx R_{\star} \frac{0.6 q_{\rm pCE}^{2/3} + {\rm ln}\left(1+q_{\rm pCE}^{1/3}\right)}{0.49 q_{\rm pCE}^{2/3}},
\label{eq:aRLOF}
\ee
where $q_{\rm pCE} \equiv M_{\star}/M_{\bullet}$ is the ratio of the donor to the accretor star mass.  The stability of the mass-transfer process depends on the response of the binary orbit and hence the donor star Roche radius to mass transfer, in comparison to the adiabatic response of the donor radius (e.g., \citealt{Soberman+97,Ge+20}). Depending on the donor's evolutionary state (in particular, the depth of its convective envelope; (e.g., \citealt{Pavlovskii+17,VignaGomez+20,Klencki+21}), dynamical instability of the binary is predicted to occur above a critical mass ratio $q_{\rm pCE} \gtrsim q_{\rm crit} \sim 0.3-3$ (e.g., \citealt{Claeys+14, Ge+15, Ge+20, vanSon+22}), though for He-rich supergiant donors much larger stability thresholds $q_{\rm crit} \gtrsim 20$ are possible (e.g., \citealt{Quast+19}).  

Assuming that mass-transfer is initiated by the post main-sequence expansion of the donor star, instability is the most likely outcome after the donor has evolved significantly to become a giant or supergiant of radius $R_{\star} \sim 1000R_{\odot}$ (though dynamical instability may be preceded or even supplanted by a phase of thermal-timescale stable mass-transfer; e.g.,  \citealt{vandenHeuvel17,Pavlovskii+17,Klencki+21,Marchant+21,vanSon+22b}).  Such highly evolved donors are comprised of a compact He core of mass $M_{\rm He} \sim 2-10M_{\odot}$ (depending on the ZAMS mass $M_{\rm ZAMS} \sim 10-30M_{\odot}$ and evolutionary state; e.g., \citealt{Woosley17}) and radius $\lesssim few \times R_{\odot}$ (e.g., \citealt{Hall&Tout14}), surrounded by a radially-extended hydrogen envelope.  

The initial spiral-in of the accretor after it loses corotation with the envelope is fast $\lesssim$ decades (e.g., \citealt{Podsiadlowski01,Ivanova+13}), but slows considerably once the companion approaches the radiative zone separating the He-rich core from the outer convective envelope, at a typical separation $R_{\rm rad} \sim 3-30R_{\odot}$ (e.g., \citealt{Passy+12,Ohlmann+16,Sand+20,Lau+22a,Hirai+Mandel22,Renzo+23}), proceeding thereafter on the longer thermal timescale (e.g., \citealt{Fragos+19,VignaGomez+22,Hirai+Mandel22}).  
During these phases, a fraction of the original envelope is unbound and ejected from the system.  The remaining material remains gravitationally bound, a portion of which is retained as an envelope surrounding the He core, while the rest resides outside the newly-formed inner binary comprised of the stripped core and the compact object (e.g., \citealt{Gotberg+20,Law-Smith+20}).   To cite a recent example from the literature, \citet{Lau+22b} found that at least $\approx 20\%$ of the envelope mass of the original $12 M_\odot$ red giant donor remains bound by the end of the hydrodynamic simulations, which is a lower bound because they do not account for radiative losses and hence may overestimate the envelope thermal energy.  

Once enough mass is removed from the He core envelope, the core will detach from its Roche surface, marking at least an temporary end to the CE phase.  However, the envelope can expand again as it re-establishes thermal equilibrium over typically hundreds to thousands of years, potentially leading to a second phase of RLOF onto the compact companion (e.g., \citealt{Dewi&Pols03,Ivanova+03,Ivanova11,Fragos+19,Romero-Shaw+20,VignaGomez+22, Laplace21}). 

Although CE events are frequently envisioned$-$if not explicitly modeled$-$as quasi-spherical 1D systems, such a structure must eventually fail to describe the system due to the substantial angular momentum imparted to the envelope during the inspiral phase (e.g., \citealt{Morris&Podsiadlowski06}), coupled with the eventual loss of radial pressure support due to radiative cooling.  Even though the simulations of \citet{Lau+22a,Lau+22b} neglect radiative losses, the residual bound envelope at the end of their simulations could arguably be best described as a thick, rotationally-supported disk surrounding the inner binary than a radial-pressure-supported envelope.  The rapid early dynamical CE phase could eventually give rise to a long-lasting {\it circumbinary disk} (CBD) phase (see also  \citealt{Kashi&Soker11,Dermine+13,Kuruwita+16,Gagnier&Pejcha22}).

\subsection{Outline of the Circumbinary Disk Evolution}
\label{sec:outline_of_disk_evolution}

\begin{figure*}
    \centering
    \includegraphics[width=1.0\textwidth]{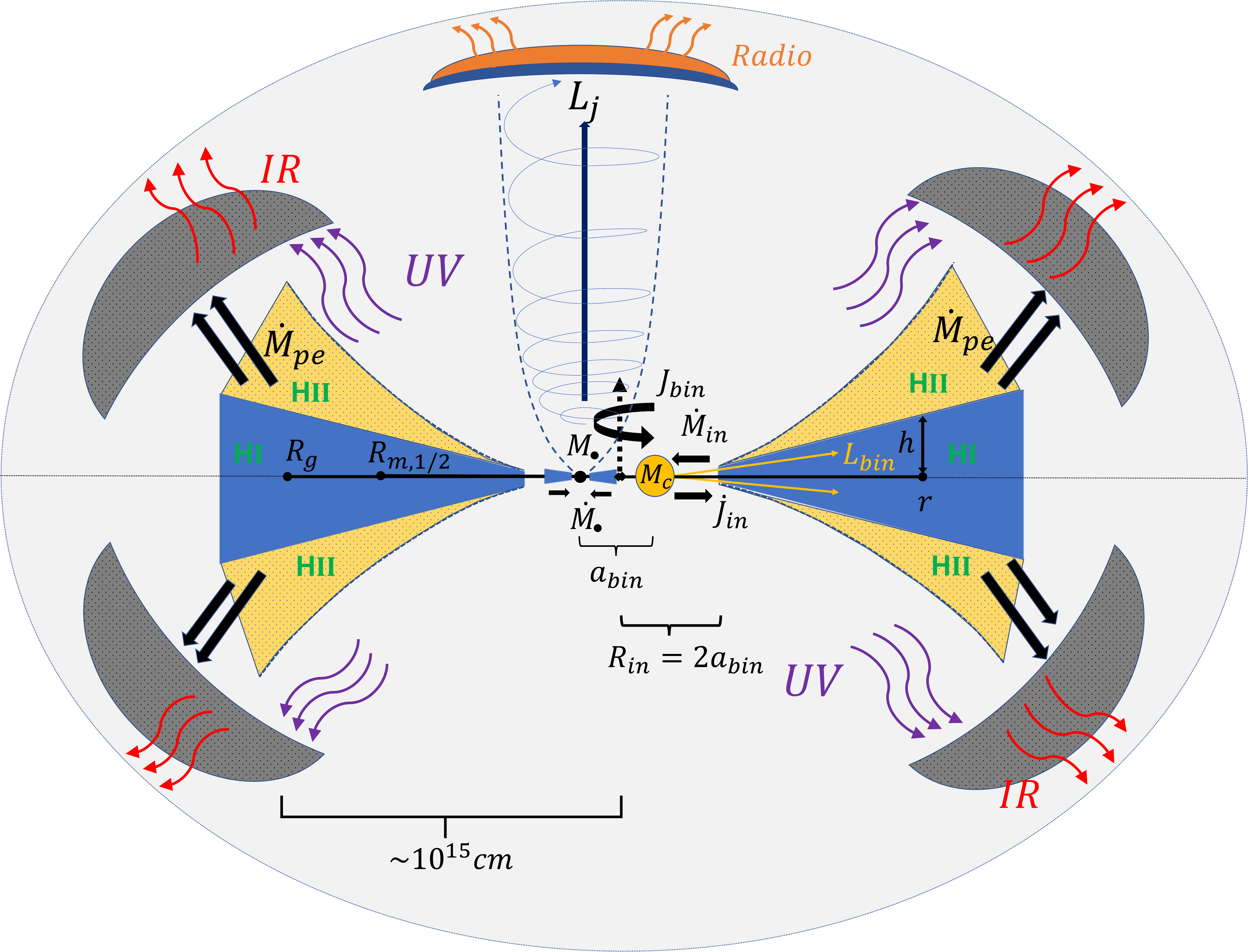}
    \caption{Schematic diagram of the post-CE system, showing the circumbinary disk surrounding the inner binary comprised of the stripped He core and BH/NS companion of separation $a_{\rm bin}$. The surface layers of the disk (shown in yellow, labelled HII) are irradiated and photo-ionized by the UV luminosity of the He core and the BH/NS disk, driving a wind from radii $\gtrsim 0.1R_{\rm g} \gtrsim 10^{14}$ cm.  During late stages of evolution, the rate of wind mass-loss becomes sufficiently high for the dense outflow to form dust (grey, crescent ``clouds"), which efficiently absorbs the binary UV luminosity, reprocessing it to near-infrared wavelengths (Sec.~\ref{sec:IR}).  CBD-fed super-Eddington accretion onto the BH/NS drives a (trans)relativistic outflow/jet, powering an parsec-scale bipolar synchrotron radio nebula (Sec.~\ref{sec:hypernebulae}).}
    \label{fig:cartoon}
\end{figure*}

We now outline the properties and evolution of the CBD, as summarized schematically in Fig.~\ref{fig:cartoon}, in order to preface the detailed model presented in the next section.  

Following the initial inspiral phase, we assume that a fraction $f_{\rm m} < 1$ of the original envelope mass resides in a CBD of initial mass
\begin{align}
M_{\rm d,0} = f_{\rm m}(M_\star - M_{\rm c}) \label{eq:fm_def},
\end{align}
where $M_{\rm c}$ is the donor star's core mass.  We consider $f_{\rm m}$ to be a free-parameter, though characteristic values $\gtrsim 10\%$ are motivated by dynamical-phase CE simulations (e.g., \citealt{Lau+22a,Lau+22b}) and the minimum accreted mass required to induce significant eccentricity growth in the post-CE binary (e.g., \citealt{Artymowicz+91,Lubow91,Siwek+23}) consistent with observed eccentricities of post-CE (e.g., \citealt{Delfosse+99}) and post-AGB (e.g., \citealt{vanWinckel18}) systems.  The disk is orbiting the inner binary of semi-major axis $a_{\rm bin} \sim 3-30 R_{\odot}$ and mass $M_{\rm bin} = M_{\bullet} + M_{\rm c}$.  We neglect any growth in the BH/NS mass during the CE phase itself (e.g., \citealt{MacLeod+17b,De+20}).

A characteristic initial radius for the gaseous disk, $R_{\rm d,0}$, can be estimated by parameterizing the specific angular momentum of the disk as a fraction $f_{\rm j}$ of the initial specific orbital angular momentum of the binary at the onset of CE, so that
\begin{align}
    \frac{f_{\rm j} q_{\rm pCE}}{(1 + q_{\rm pCE})^2}\sqrt{G(M_\star + M_\bullet) a_{\rm RLOF}} \approx \sqrt{G(M_{\rm c} + M_\bullet) R_{\rm d,0}}, \label{eq:fj_def}
\end{align}
where again $q_{\rm pCE} \equiv M_\star/M_\bullet$ is the mass-ratio of the pre-CE binary. This yields
\begin{align}
    R_{\rm d,0} = a_{\rm RLOF}f_{\rm j}^2\frac{q_{\rm pCE}^2}{(1 + q_{\rm pCE})^4}\frac{M_{\star}+M_{\bullet}}{M_{\rm bin}}.
    \label{eq:Rd0}
\end{align}
For example, for $M_{\bullet} = 10M_{\odot}$, $M_{\rm c} = 10 M_{\odot}$, $M_{\star} = 25M_{\odot}$, $R_{\star} = 1000 R_{\odot}, M_{\rm bin} = 20M_{\odot}, q_{\rm pCE} = 2.5$, $f_{\rm j} = 1$ we find $a_{\rm RLOF} \approx 2200 R_{\odot}$ and $R_{\rm d,0} \approx 160 R_{\odot}.$  The characteristic initial surface density of the disk can then be estimated as
\begin{align}
    \Sigma_{\rm 0} &\sim \frac{M_{\rm d,0}}{2\pi R_{\rm d,0}^2} \approx 1.7\,\times\,10^{6}\,\text{g}\,\text{cm}^{-2}\left(\frac{M_{\rm d,0}}{ M_\odot}\right)\left(\frac{R_{\rm d,0}}{200 R_\odot}\right)^{-2}.
    \label{eq:Sigma0}
\end{align}
These disk properties are broadly consistent with \citet{Lau+22a,Lau+22b} who find by the end of their CE simulations a bound envelope arranged in a flattened disk-like configuration containing a few solar masses with a radial density profile $\rho \propto r^{p}$ with $p \approx -2$ on radial scales $r \gtrsim 10^2-10^{3}R_\odot$ (M. Lau, private communication); we note that for this density profile, the disk's total mass $\propto \int \rho r^{2}dr$ is indeed dominated by large radii $\sim R_{\rm d,0}$.  Matching to a characteristic disk radius $R_{\rm d,0} \sim 10^{2}-10^{3}R_{\odot}$ for the binary parameters adopted by \citet{Lau+22a,Lau+22b} would imply $f_{\rm j} \approx 1-3$. 

The initial phases of the CE phase are so rapid as to be effectively adiabatic, rendering the CBD initially hot and thick, with a vertical aspect ratio $\theta \equiv h/r \sim 1$, where $h \simeq c_{\rm s}/\Omega$ is the vertical scale-height in hydrostatic equilibrium, $c_{\rm s}$ is the midplane sound speed and $\Omega = \Omega_{\rm K} \equiv (GM_{\rm bin}/r^{3})^{1/2}$ is the angular velocity (approximated as Keplerian, i.e. neglecting the disk's self-gravity).  Radiation pressure greatly exceeds gas pressure in the disk, at least initially.  The disk thus cools radiatively on the vertical photon diffusion timescale,
\begin{align}
&t_{\rm cool} \approx \tau\frac{h}{c}\nonumber\\
& \approx 10\,{\rm yr}\, \left(\frac{\theta}{0.2}\right)\left(\frac{\kappa}{3\,{\rm cm^{2}\,g^{-1}}}\right)\left(\frac{M_{\rm d,0}}{M_\odot}\right)\left(\frac{R_{\rm d,0}}{200 R_\odot}\right)^{-1},
\label{eq:tcool}
\end{align}
where $\tau = \kappa \Sigma_0/2$ is the vertical optical depth through the disk midplane, $\kappa$ is the opacity, and the values of $\theta$ and $\kappa$ are normalized to values typical of those achieved in hydrostatic and thermal balance for fiducial values of $\Sigma_{\rm d,0}$ and $R_{\rm d,0}$ (Eqs.~\eqref{eq:Rd0}, \eqref{eq:Sigma0}).  The cooling time being only a few orders of magnitude longer than the dynamical time 
\be
t_{\rm dyn} \sim \frac{1}{\Omega} \approx 0.03\,{\rm yr}\,\left(\frac{R_{\rm d,0}}{200\,R_{\odot}}\right)^{3/2}\left(\frac{M_{\rm bin}}{20M_{\odot}}\right)^{-1/2},
\ee
demonstrates that radiative losses can become important relatively soon after the dynamical plunge (e.g., \citealt{Fragos+19}; however, see \citealt{Gagnier&Pejcha22}).

\begin{figure*}
    \centering
    \includegraphics[width=1.0\textwidth]{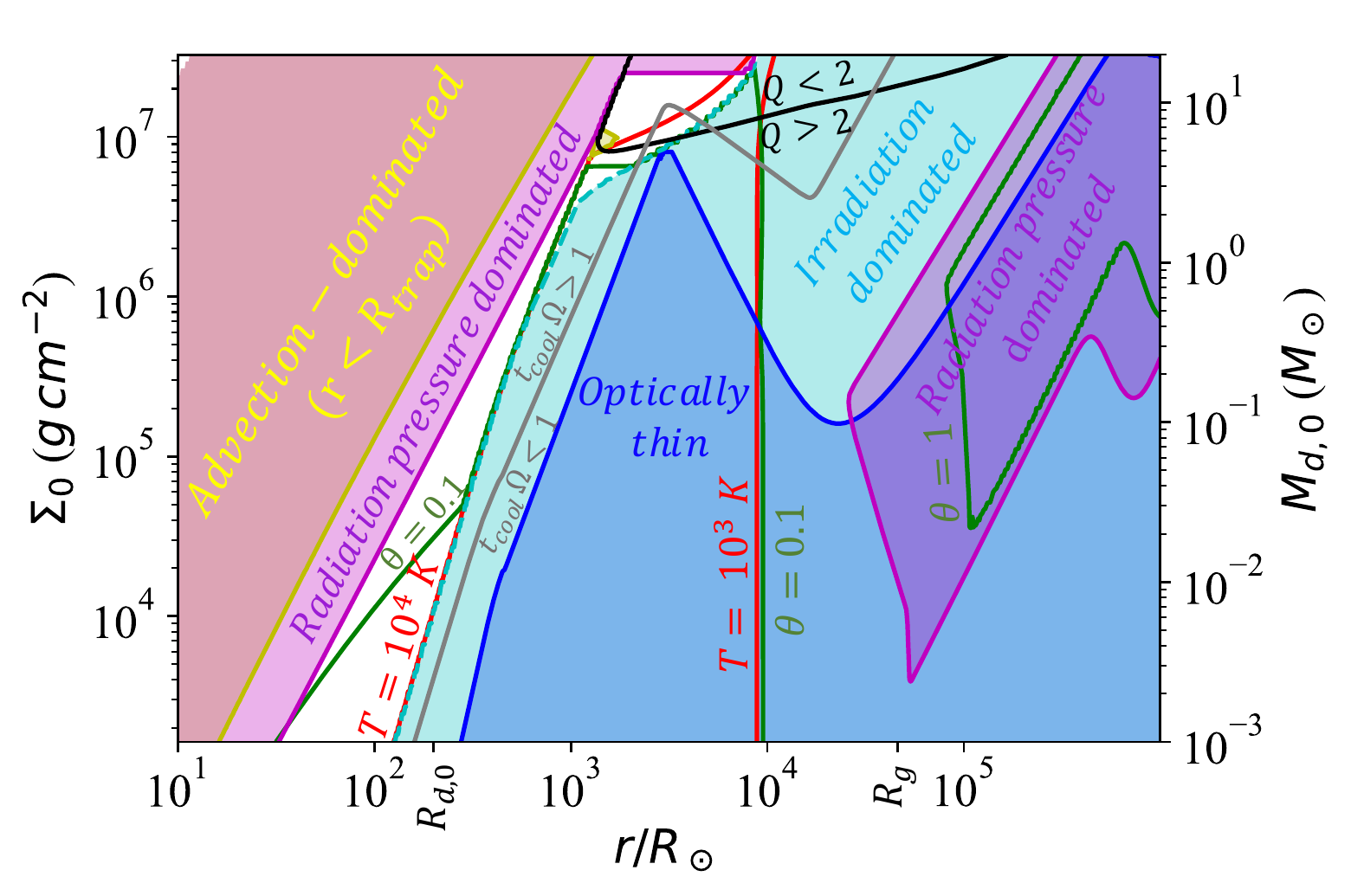}
    \caption{Physical properties of post-CE CBD in the space of local mass $M_{\rm d} \equiv 2\pi\,r^2\Sigma$ and radius, calculated from vertical hydrostatic and thermal balance (Sec.~\ref{sec:model}).  We adopt a density profile $\Sigma = \Sigma_0 \left(r/R_{\rm d,0}\right)^{-2}$, for an assumed $R_{\rm d,0} = 200R_{\odot}$ motivated by Eq.~\eqref{eq:Rd0} and normalization $\Sigma_{0} \equiv \Sigma(R_{\rm d,0})$ related to the total disk mass $M_{\rm d,0}$ according to Eq.~\eqref{eq:Sigma0}.  We otherwise adopt the parameters of the fiducial model: binary mass $M_{\rm bin} = 20 M_\odot$, luminosity $L_{\rm bin} = L_{\rm Edd}$, viscosity $\alpha = 0.01$, advective cooling parameter $\xi = 1$ (Table~\ref{tab:parameters_symbols}).  We mark the characteristic initial disk radius $R_{\rm d,0}$ and the radius $R_{\rm g}$ (Eq.~\eqref{eq:Rg}) external to which photoevaporation mass-loss becomes important as ticks along the bottom axis. }
    \label{fig:paramspace}
\end{figure*}

As the disk cools from its initial state on timescales $\gtrsim t_{\rm cool}$, a balance will be reached between heating and radiative cooling, eventually resulting in a thinner equilibrium disk structure $\theta \ll 1$.  Prefacing our detailed model results in Sec.~\ref{sec:model}, Figure \ref{fig:paramspace} summarizes the regimes of disk properties in thermal and vertical hydrostatic balance in the space of local disk mass $\propto 2\pi r^{2} \Sigma$ and radius $r$.  Green contours show that an equilibrium vertical thickness $\theta \sim 0.1$ is achieved across a wide range of radii and masses.

As in the outer regions of protostellar disks, a key source of disk heating is provided by the UV luminosity of the central binary (e.g., \citealt{Chiang&Goldreich97}); the latter is comprised of the He core nuclear luminosity and the accretion luminosity onto the BH/NS, both of which are expected to radiate close to their respective Eddington luminosities, giving a combined luminosity $L_{\rm bin} \sim L_{\rm Edd} \approx 3\times 10^{39}(M_{\rm bin}/20M_{\odot})$ erg s$^{-1}$ (we show below that the accretion rate onto the central binary exceeds the Eddington rate).  Also as in protostellar disks, while ``viscous'' heating from turbulent angular momentum transport dominates at small radii in the disk $\lesssim 10^{2}R_{\odot}$, irradiation heating dominates further out (the teal contour in Fig.~\ref{fig:paramspace} delineates this boundary; see also Appendix \ref{app:analytic}). 

The disk is optically thin across a range of radii $r \sim 10^{3}R_{\odot}$ (blue region marked $\tau < 1$ in Fig.~\ref{fig:paramspace}).  The disk midplane temperatures $T \approx 10^3 - 10^{4}$ K in this region (between the red contours in Fig.~\ref{fig:paramspace}) correspond to the ``gap'' between which atomic and molecular processes dominate the opacity.  This necessitates including a detailed opacity law in our model (Sec.~\ref{sec:model}), again similar to protostellar disks and the outer regions of AGN disks.  

As the gaseous disk cools to a thinner configuration, torques from the inner binary will eventually truncate the disk outside the binary orbit near a radius \citep{MacFadyen&Milosavljevic08,Miranda&Lai15}
\begin{align}
    R_{\rm in} \approx 2 a_{\rm bin} \sim 10-100R_{\odot} \label{eq:Rin}.
\end{align}
Gas which accretes through $R_{\rm in}$ feeds matter onto the central binary (e.g., \citealt{Farris+15,Ryan&MacFadyen17}), partitioned into separate ``mini-disks'' orbiting the BH/NS and the stellar core, respectively (assuming the latter has become sufficiently detached from its Roche lobe to fit a disk around it).
 
Various sources of turbulence in the disk, such as thermal convection \citep{Gagnier&Pejcha22} and the magneto-rotational instability (MRI; \citealt{Balbus&Hawley98}), remove angular momentum from the gas, allowing it to accrete inwards on the viscous timescale,
\begin{align}
    &t_\nu = \frac{r^{2}}{\nu} \approx  3.2\,\times\,10^2\,\text{yr}\times\nonumber\\
    &\left(\frac{\alpha}{0.01}\right)^{-1}\left(\frac{M_{\rm bin}}{20 M_\odot}\right)^{-1/2}\left(\frac{r}{200 R_\odot}\right)^{3/2}\left(\frac{\theta}{0.1}\right)^{-2}
    \label{eq:tnu}
\end{align}
where $\nu = \alpha r^{2}\Omega \theta^{2}$ is the kinematic viscosity and $\alpha \sim 10^{-2}$ is the \citet{Shakura&Sunyaev73} viscosity parameter (Eq.~\eqref{eq:nu}).  The resulting initial accretion rate,
\begin{align}
&\dot{M} \sim \frac{M_{\rm d,0}}{t_{\nu}(R_{\rm d,0})} \approx 3\times 10^{-3}M_{\odot}{\rm yr^{-1}}\,\times\nonumber\\
&\left(\frac{\alpha}{0.01}\right)\left(\frac{M_{\rm d,0}}{M_{\odot}}\right)\left(\frac{M_{\rm bin}}{20 M_\odot}\right)^{1/2}\left(\frac{R_{\rm d,0}}{200 R_\odot}\right)^{-3/2}\left(\frac{\theta}{0.1}\right)^{2},
\end{align}
exceeds by several orders of magnitude the BH/NS Eddington accretion rate, $\dot{M}_{\rm Edd} \equiv L_{\rm Edd}/(0.1 c^{2}) \simeq 3\times 10^{-7}(M_{\bullet}/10M_{\odot})M_{\odot}$ yr$^{-1}$.  On the other hand, the ``trapping radius'' \citep{Begelman79} 
\be
R_{\rm tr} \approx 10R_{\rm G}(\dot{M}/\dot{M}_{\rm Edd}) \lesssim R_{\odot}, 
\label{eq:Rtrap}
\ee
inside of which photons are advected radially inwards by the accretion flow faster than they can diffuse outwards, resides near or inside the binary orbit and Roche surface of the BH/NS, where $R_{\rm G} \equiv GM_{\bullet}/c^{2}$.  Photon trapping and advective cooling are thus only important within the CBD itself at the earliest times.  

Tidal interactions exchange angular momentum between the gas disk and binary \citep{Lin+Papalouizou79, Goldreich+Tremaine80, Pringle91}.  The binary mass ratio is sufficiently large ($M_{\bullet}/M_{\rm c} \sim 0.1-1$) to open a gap in the disk (e.g., \citealt{Artymowicz&Lubow94,Lai&Munoz22}).  A growing body of work explores the magnitude and sign of the torque, and how these vary with the disk properties (e.g., \citealt{Duffell+20,Munoz+20,Gagnier&Pejcha22,Siwek+23}; see Sec.~\ref{sec:binaryevo}).  To order of magnitude, the binary semi-major axis is increased or decreased by gas accretion on a timescale (e.g., \citealt{Haiman+09})
\begin{align}
t_{\rm \dot{a}, \rm CBD} &\equiv \frac{a_{\rm bin}}{\dot{a}_{\rm bin}} \sim \frac{\mu_{\rm bin}}{\dot{M}_{\rm bin}}\nonumber\\
&\sim \frac{\mu_{\rm bin}}{M_{\rm d,0}}t_{\nu}(R_{\rm d,0}) \sim 10^{3}-10^{4}\,{\rm yr},
\label{eq:tadot}
\end{align}
where $J_{\rm bin} = \mu_{\rm bin}a_{\rm bin}^{2}\Omega_{\rm bin}$ is the binary angular momentum, $\mu_{\rm bin} \equiv M_{\bullet}M_{\rm c}/M_{\rm bin}$, and the estimate in the final line follows from Eq.~\eqref{eq:tnu} for $M_{\rm d,0}/\mu_{\rm bin} \sim \theta \sim 0.1$ (Fig.~\ref{fig:paramspace}). 

Following envelope removal during the CE phase, the He core will re-establish thermal equilibrium.  This occurs on roughly the Kelvin-Helmholtz timescale,
\begin{align}
    & t_{\rm KH} = \frac{GM_{\rm c}^2}{2R_{\rm c}L_{\rm c}}\nonumber\\
    &\approx 8\times 10^3\,{\rm yr}\,\left(\frac{M_{\rm c}}{10\,M_\odot}\right)\left(\frac{R_{\rm c}}{R_\odot}\right)^{-1}\left(\frac{L_{\rm c}}{0.5L_{\rm edd,c}}\right)^{-1}, \label{eq:t_kh}
\end{align}
where the He core luminosity $L_{\rm c}$ and radius $R_{\rm c}$ are normalized to characteristic values \citep{Gotberg+19} and $L_{\rm Edd,c}$ is the Eddington luminosity of the core.  The nuclear evolution timescale of the core is generally longer, varying from $10^{4}$ to $10^5\,\rm yr$ (e.g., \citealt{Meynet+94}), depending on its evolutionary state at the time of the original CE.  A delay until core collapse as short as thousands of years is possible if the mass transfer commenced after core He depletion (so called case C RLOF, see \citealt{Lauterborn1970}) a circumstance more likely to occur at low metallicity (see \citealt{Klencki+20}, their Fig.~3).

In conclusion, we arrive at a rough hierarchy of timescales in the post-CE binary,
\be
t_{\rm dyn} \ll t_{\rm cool} \ll t_{\dot{a}, \rm CBD} \sim t_{\rm KH} \lesssim t_{\rm nuc},
\ee
This demonstrates that any post-CE accretion phase could have a significant impact on the inner binary properties, prior to the core-collapse or even thermal re-expansion of the core. 

\section{Model}
\label{sec:model}

We model the coupled evolution of the inner binary and the surrounding gaseous CBD.  We begin by describing the evolution equations for the disk (Sec.~\ref{sec:diskevo}) and binary (Sec.~\ref{sec:binaryevo}) before discussing technical details and boundary conditions associated with the numerical calculation (Sec.~\ref{sec:numerical}).

\subsection{Disk Evolution}
\label{sec:diskevo}

The time evolution of the surface density $\Sigma(r,t)$ of a thin, axisymmetric, Keplerian disk is governed by the mass and angular momentum conservation equations, which together lead to \citep{Frank+02, armitage_2020}
\be
\frac{\partial\Sigma}{\partial{t}} = \frac{3}{r}\frac{\partial}{\partial{r}}\left(r^{1/2}\frac{\partial}{\partial{r}}(\nu\Sigma{r^{1/2}})\right) - \dot{\Sigma}_{\rm pe},
\label{eq:sigma_diff_eq}
\ee
where $\dot{\Sigma}_{\rm pe}$ accounts for mass-loss due to photoevaporation by the central binary (Sec.~\ref{sec:photoionization}).  As described in the next section, we model angular momentum transfer between the CBD and the binary by injecting angular momentum into the disk at the inner boundary; angular momentum sink terms therefore do not appear explicitly in Eq.~\eqref{eq:sigma_diff_eq}.  

The kinematic viscosity \citep{Shakura&Sunyaev73}
\begin{align}
    \nu(r, \Sigma) = \alpha\frac{P}{\rho\Omega} = \alpha r^{2}\Omega \theta^{2}, \label{eq:nu}
\end{align}
is assumed to scale with the total midplane pressure $P$, where $\rho = \Sigma/2h$ is midplane density.  Values of $\alpha \sim 10^{-2}-10^{-1}$ are motivated by MRI simulations (e.g., \citealt{Davis+10}) and observations (e.g., \citealt{King+07}).  The MRI may be suppressed in neutral regions of the disk (e.g., \citealt{Gammie96,Gammie&Menou98}); however, the midplane is either sufficiently hot to be thermally-ionized ($T \gtrsim 10^{3}$ K; Fig.~\ref{fig:paramspace}), or optically-thin to be photo-ionized, at all times and radii we simulate for the MRI to be active. 

\subsubsection{Disk Vertical Structure}
\label{sec:disk_vertical_structure}

The disk scale-height and midplane pressure are determined, respectively, by vertical hydrostatic balance and thermal equilibrium,
\begin{align}
P &= \rho h^2 \Omega^2 \label{eq:hydrostatic_eq}\\
F_{\rm visc} &= \frac{1}{f(\tau)}\left(2\sigma T^4 - F_{\rm irr}\right) + F_{\rm adv}\label{eq:thermal_balance}
\end{align}
where $T$ is the midplane temperature and 
\be
F_{\rm visc} = \frac{9}{4}\nu\Omega^2\Sigma,
\ee
is the viscous heating rate per unit area \citep{Frank+02}, where $f(\tau)$ determined by vertical radiation transfer \citep{Sirko+Goodman03,kratter2010runts,Margalit&Metzger16}
\begin{align}
    f(\tau) = \frac{3}{2}\tau + 2 + \frac{1}{(\tau + \tau_{\rm min})},
\end{align}
$\tau = \kappa \Sigma/2$, and $\kappa(\rho,T)$ is the Rosseland mean opacity.  The cut-off $\tau_{\rm min}$ is included to avoid overestimating the midplane temperature due to the breakdown of the LTE assumption in the optically-thin limit $\tau \rightarrow 0$ \citep{Hubeny90}.  We set $\tau_{\rm min} = 0.01$, though we have checked that our results are not sensitive to this choice as long as $\tau_{\rm min}$ is small but non-zero.

At small radii and early times, photons are trapped in the disk, which becomes radiatively inefficient and geometrically thick $\theta \sim \mathcal{O}(1)$, such that radial advection of heat becomes faster than radiative cooling.  We account for this by including an advective cooling term of the form (e.g., \citealt{Narayan&Yi95})
\begin{align}
    F_{\rm adv} &= T|v_r|\frac{ds}{dr}
\end{align}
where $v_r$ is the radial velocity, and $s = s_{\rm gas} + s_{\rm rad}$ is the specific entropy per unit surface area, comprised of gas and radiation contributions.  Instead of calculating this term self-consistently using $v_r$ and $ds/dr$ derived from radial derivatives of $\Sigma$, we make the approximation (e.g., \citealt{DiMatteo+02})
\begin{align}
    F_{\rm adv} &\approx \xi\frac{3\nu}{2r}\theta\,T\frac{s_{\rm gas} + s_{\rm rad}}{r} = \xi\frac{3\nu}{2r}\theta\,\left(\frac{3}{2}P_{\rm gas} + 4P_{\rm rad}\right), \label{eq:cooling_adv}
\end{align}
where the dimensionless parameter $\xi$ is typically of order unity, since $v_r \approx -3\nu/2r$, $ds/dr = (d\log{s}/d\log{r})s/r$ and typically, $|d\log{s}/d\log{r}| \sim \mathcal{O}(1)$.  We set $\xi = 1$ fiducially but explore the sensitivity of our results for the disk evolution on its value.  

As described in Sec.~\ref{sec:diskevo}, we tabulate solutions of the vertical structure as a function of $\Sigma$ and $r$ which we interpolate between to obtain $\nu(\Sigma,r)$ to drive the time evolution.  In regions of the disk where advective cooling dominates and $\theta \sim 1$, radial pressure gradients also become important and the angular velocity can be sub-Keplerian $\Omega < \Omega_{\rm K}$; however, we ignore these corrections and assume $\Omega = \Omega_{\rm K}$ everywhere (see also \citealt{DiMatteo+02}).

We adopt a Rosseland mean opacity law from \citet{Bell&Lin94}, which interpolates opacities of the form $\kappa \propto \rho^i\,T^j$ in 8 different regions in the temperature-density plane with corresponding power law indices. This opacity law assumes solar metallicity and includes contributions from ice/metal grains at very low temperatures, molecular opacity and H-scattering at mid-temperatures and bound-free, bound-bound and electron scattering.  

We approximate the flux of radiation from the binary on the disk surface at a given radius by
\be
F_{\rm irr} = \frac{L_{\rm bin}}{2\pi R^2}\theta
\ee
where 
\begin{align}
    L_{\rm bin} = \eta_{\rm Edd}L_\text{Edd} \approx 2.8\times 10^{39}\,\text{erg}\,\text{s}^{-1}\eta_{\rm Edd}\left(\frac{M_{\rm bin}}{20 M_\odot}\right)
    \label{eq:Lbin}
\end{align}
is the binary luminosity and $\eta_{\rm Edd} \lesssim 1$.  We ignore order-unity factors and other complications such as self-shielding that arise from radial-dependence of the disk flaring angle \citep{Bell+97} and associated thermal instabilities \citep{Watanabe&Lin08}, instead exploring the sensitivity of our results to $\eta_{\rm Edd}$. 

The equation of state is an ideal gas of particles and radiation in local thermal equilibrium, with total pressure
\be
P = P_{\rm gas} + P_{\rm rad},
\ee
where $P_{\rm gas} = k\rho T/\mu m_p$ is the gas pressure, $k$ is the Boltzmann constant, $\mu$ is mean molecular weight, and $m_p$ is the proton mass.  Depending on the ionization state and composition of the disk material (e.g., atomic vs.~molecular), the mean weight can vary from $0.68 \lesssim \mu \lesssim 2.3$; for simplicity we set $\mu = 2$ throughout our simulations, but we have checked that our results do not depend sensitively on this assumption.

In optically-thick regions of the disk where $\tau \gg 1$, the radiation pressure is given by $P_{\rm rad} = aT^4/3$, where $a$ is the radiation constant.  In optically-thin regions, the outwards radiative flux and its probability of interacting with the gas are each attenuated by a factor of $\tau$ \citep{Sirko+Goodman03}, such that $P_{\rm rad} \propto \tau^2 T^4$. We interpolate between these two limits according to:
\begin{align}
    P_{\rm rad} = (1 - e^{-\tau^2})\frac{aT^4}{3} \label{eq:rad_pressure}.
\end{align}
\subsubsection{Gravitational instability}
\label{sec:grav_instability}
For sufficiently massive disks, solutions of the vertical structure equations outlined in Sec.~\ref{sec:disk_vertical_structure} are unstable to radial perturbations driven by self-gravity (\citealt{Toomre64, Paczynski78,Kratter&Lodato16}). The instability arises for the values of the Toomre Q parameter \citep{Toomre64} which satisfy
\begin{align}
    Q = \frac{c_s\Omega}{\pi\,G\Sigma} = \frac{\Omega^2}{\pi\,G\rho} \leq Q_0 ,
    \label{eq:Toomre}
\end{align}
where $Q_0 \approx 2.$  When the gravitational instability sets in, spiral-arm waves are generated in a dynamical timescale throughout the disk, leading to a gravito-turbulent state (e.g., \citealt{Gammie01,Lodato&Rice04}).  The outcome of this instability depends on the local thermal state of the disk.  If the vertical cooling time of the disk is long compared to the dynamical time ($t_{\rm cool} \Omega \gg 1$), the turbulence saturates in a marginally unstable state with $Q \approx Q_0$, due to a self-regulating feedback loop \citep{Gammie01}.  By contrast, if cooling is efficient ($t_{\rm cool} \Omega \ll 1$), the disk fragments into smaller, gravitationally-bound objects \citep{Gammie01}.  

In practice we find that the instability condition is met only across a limited region of parameter space in our models, for high disk masses at intermediate radii within the opacity gap.
When it arises we handle it by replacing the thermal balance equation (Eq.\eqref{eq:thermal_balance}) with the marginal stability condition,
\begin{align}
    Q = Q_0 = 2,
\end{align}
wherever $Q \leq Q_0$ (calculated neglecting self-gravity).  We neglect the possible impact of fragmentation, which can occur when $t_{\rm cool}\Omega > 1$, and its possibly interesting implications for planet formation in this exotic environment (e.g., \citealt{Perets10,Schleicher&Schleicher14}).  

\subsubsection{Photoionization mass-loss}
\label{sec:photoionization}

Photo-evaporation by the EUV radiation field from the central binary becomes important at radii approaching \citep{Hollenbach+94}
 \be R_{\rm g} \equiv \frac{GM_{\rm bin}}{c_{\rm s,0}^{2}} \approx 2\times 10^{15}\,{\rm cm}\,\left(\frac{M_{\rm bin}}{20M_{\odot}}\right)\left(\frac{\mu_{\rm pe}}{0.68}\right),
 \label{eq:Rg}
 \ee
where $c_{\rm s,0} \approx (kT_{\rm 0}/\mu m_p)^{1/2}$, $T_{\rm 0} \sim 10^{4}$ K, and $\mu_{\rm pe} \approx 0.68$ are the sound speed, temperature, and mean molecular weight of the photoionized upper layers of the disk.  For the sink term in Eq.~\eqref{eq:sigma_diff_eq}, we take \citep{Hollenbach+94,Adams+04,Anderson+13}
  \be \dot{\Sigma}_{\rm pe} = \frac{\rho_0 c_{\rm s,0}}{4\pi}\left(\frac{R_{\rm g}}{r}\right)^{3/2}\left[1+\frac{R_{\rm g}}{r}\right]\exp\left[-\frac{R_{\rm g}}{2r}\right] \label{eq:sigma_dot_pe},
 \ee 
 where the density at the base of the outflow $\rho_0 = {\rm min}\left[\rho, \rho_{\rm pe}\right]$ is given by the minimum of the midplane density and that of the photoionized layer $\rho_{\rm pe}$ set by the location where photoionization and recombination balance.  For the latter we follow \citet{Hollenbach+94} (their Eq.~3.11) 
 \begin{eqnarray} \rho_{\rm pe} &\simeq& 3\times 10^{-17}\,{\rm g\,cm^{-3}}\,\left(\frac{\Phi}{10^{49}\,\rm s^{-1}}\right)^{1/2}\left(\frac{r}{10^{15}\,{\rm cm}}\right)^{-3/2} \nonumber \\
 &\approx& 2\times 10^{-17}\,{\rm g\,cm^{-3}}\, \left(\frac{L_{\rm bin}}{L_{\rm Edd}}\right)^{1/2}\left(\frac{M_{\rm bin}}{20M_{\odot}}\right)^{-1}\left(\frac{r}{R_{\rm g}}\right)^{-3/2}, \label{eq:rho_pe}
 \end{eqnarray}
 where in the second line we have estimated the number of ionizing photons emitted per second from the binary as $\Phi \simeq L_{\rm bin}/(2\,{\rm Ry})$, where Ry $= 13.6$ eV, motivated by the high effective temperatures of both the stripped core ($T_{\rm eff} \approx 10^{5}$ K; e.g., \citealt{Gotberg+19}) and the BH/NS accretion disk (e.g., \citealt{Kovlakas+22}).  The characteristic maximum mass-loss rate at radii $r \sim R_{\rm g}$ (assuming $\rho_0 = \rho_{\rm pe} < \rho$) is given by
 \begin{align}
 \dot{M}_{\rm pe} &\simeq 2\pi \dot{\Sigma}_{\rm pe} R_{\rm g}^{2} \sim \rho_{\rm pe}(R_{\rm g})c_{\rm s,0}R_{\rm g}^{2}\nonumber\\
 &\approx 10^{-5}M_{\odot}\,{\rm yr^{-1}}\,\left(\frac{L_{\rm bin}}{L_{\rm Edd}}\right)^{1/2}\left(\frac{M_{\rm bin}}{20M_{\odot}}\right).
 \end{align}

\subsection{Binary Evolution}
\label{sec:binaryevo}

We assume the inner binary is circular with a constant mass ratio $q = M_{\rm c}/M_{\bullet}$ at all times (despite the fact that the CBD is likely to impart moderate eccentricity to the binary; Sec. \ref{sec:alpha}).  The binary orbital angular momentum is
\begin{align}
    J_{\rm bin} &= \mu_{\rm bin}a_{\rm bin}^2\Omega_{\rm bin}  = \frac{q}{(1+q)^{2}}G^{1/2}M_{\rm bin}^{3/2}a_{\rm bin}^{1/2},
    \label{eq:Jbin}
\end{align}
where $\Omega_{\rm bin} = \left(GM_{\rm bin}/a_{\rm bin}^{3}\right)^{1/2}$ and $\mu_{\rm bin} = M_{\bullet}M_{\rm c}/M_{\rm bin}$.  Changes in the binary mass or angular momentum thus change its semi-major axis according to
\begin{align}
\frac{\dot{a}_{\rm bin}}{a_{\rm bin}} = 
    2\frac{\dot{J}_{\rm bin}}{J_{\rm bin}} - 3\frac{\dot{M}_{\rm bin}}{M_{\rm bin}}
    \label{eq:adotbin1}
\end{align}
We focus exclusively on the exchange of mass and angular momentum with the CBD, as described below.  We neglect contributions to $\dot{M}_{\rm bin}$ arising from stellar wind mass-loss (e.g., \citealt{Soberman+97, Neijssel+21}), or to $\dot{M}_{\rm bin}$ or $\dot{J}_{\rm bin}$ as a result of mass-transfer between the core and compact object (we explore the effect of the CBD on the latter in Sec.~\ref{sec:stability}).  

Angular momentum exchange between the disk and the binary occurs through three channels: Lindblad resonances, corotation torques, and torques arising from thermal effects of gas near the companion orbit (e.g., \citealt{Lin+Papalouizou79,Goldreich+Tremaine80}; see \citealt{armitage_2020,Lai&Munoz22} for reviews). The interplay between these effects is complex, with ongoing debate regarding how the net torque depends on the disk and binary properties.  The binary orbital evolution is conveniently characterized by the \emph{accretion eigenvalue} \citep{Lai&Munoz22},
\begin{align}
    l_0 \equiv \frac{\dot{J}_{\rm bin}}{\dot{M}_{\rm bin}} = \chi a_{\rm bin}^{2}\Omega_{\rm bin} \label{eq:acc_eigenvalue},
\end{align}
which represents the torque on the binary per unit accreted mass, where $\chi$ is a dimensionless parameter.  In terms of this definition, Eq.~\eqref{eq:adotbin1} becomes
\begin{align}
\frac{\dot{a}_{\rm bin}}{a_{\rm bin}} =  \left[2\chi\frac{(q+1)^{2}}{q} - 3 \right]\frac{\dot{M}_{\rm bin}}{M_{\rm bin}},
    \label{eq:adotbin}
\end{align}
thus motivating the radial migration timescale $t_{\dot{a}, \rm CBD}$ estimated earlier (Eq.~\eqref{eq:tadot}).  A critical value
\be
\chi_{\rm crit} = \frac{3}{2}\frac{q}{(1+q)^{2}} \underset{q=1} \simeq 0.38
\label{eq:chi_crit}
\ee
separates binaries that widen due to mass accretion ($\dot{a}_{\rm bin} > 0; \chi > \chi_{\rm crit}$) from those that contract ($\dot{a}_{\rm bin} < 0; \chi < \chi_{\rm crit}$).

For circular $q = 1$ binaries with vertical aspect ratio $\theta = 0.1$, locally-isothermal $\alpha$-viscosity disk simulations find $\chi \simeq 0.68$ (\citealt{Munoz+20}; see also \citealt{Miranda+17,Moody+19,Duffell+20,Dittmann&Ryan21,DOrazio&Duffell21,Tiede+20,Zrake+21,Franchini+22}), giving $\dot{a}_{\rm bin} > 0$ and hence outwards binary migration.  Smaller values of $\chi < \chi_{\rm crit}$ corresponding to inwards migration are instead found for thinner disks (e.g., a threshold $\theta < 0.04$ was found by \citealt{Tiede+20}), and for the much thicker $\theta \sim 1$ disk that characterizes the earliest post-CE phase \citep{Gagnier&Pejcha22}.  On the other hand, for moderately eccentric and/or lower mass-ratio binaries $q \lesssim 0.6$, \citet{Siwek+23} found inward migration.  Even if the binary starts circular, the eccentricity is expected to quickly grow, towards an equilibrium value $e \sim 0.1-0.5$ (e.g., \citealt{Zrake+21,Siwek+23}), at which the value of $\chi$ will in general also differ from the assumed circular case.  Almost all studies to date model angular momentum transport using a hydrodynamic $\alpha$-viscosity, rather than self-consistent magnetohydrodynamic simulations (however, see \citealt{Shi+12}).  

In light of these uncertainties, we take $\chi$ to be a fixed free parameter of order unity independent of the other disk properties, and explore how the disk evolution changes for different assumed values of $\chi$.  This unfortunately precludes making definitive predictions about whether the CBD will soften or harden the binary relative to its immediate post-CE state, or capturing the potentially nontrivial behavior (e.g., switches between inwards and outwards migration) that could in principle develop for a time-dependent $\chi$, as the disk evolves from being thick to thin as it cools and spreads and the binary eccentricity evolves.  

The radial distribution of the binary torque density is sharply peaked near the inner boundary of the disk; hence we model it as a delta function at the inner boundary, reducing the interaction terms in the disk evolution to a boundary condition imposed at $r = R_{\rm in}$ \citep{Pringle91}.  The net rate of change of the disk's angular momentum near the inner boundary $\dot{J}_{\rm in} = -\dot{J}_{\rm bin}$ can be written \citep{Nixon+Pringle21}
\begin{align}
 \dot{J}_{\rm in} = \dot{J}_{\rm visc, in} + \dot{J}_{\rm adv, in} \label{eq:jdot_in},
\end{align}
where 
\begin{align}
    \dot{J}_{\rm visc,in} = 3\pi\nu\Sigma\left(GM_{\rm bin}r\right)^{1/2}, \quad r = R_{\rm in}
\end{align}
is the local viscous torque (which acts to increase the disk angular momentum) and
\begin{align}
    \dot{J}_{\rm adv,in} = \dot{M}_{\rm in}(GM_{\rm bin}R_{\rm in})^{1/2}, \quad r = R_{\rm in}
\end{align}
is the angular momentum advected through the inner boundary (which acts to decrease the disk angular momentum), where $\dot{M}_{\rm in} = -\dot{M}_{\rm bin} < 0$ is the binary accretion rate. The standard ``no-torque'' boundary condition is formulated by
\begin{align}
\dot{J}_{\rm in} = 0 \iff \dot{J}_{\rm adv, in} = -\dot{J}_{\rm visc, in}
\end{align}
When an external torque is present, from non-axisymmetric gravitational forces driven by the binary motion in our case, this no longer holds. Noting that the accretion eigenvalue $l_0$ (Eq.~\eqref{eq:acc_eigenvalue}) can be written
\begin{align}
    l_0 = \frac{\dot{J}_{\rm bin}}{\dot{M}_{\rm bin}} \approx \frac{\dot{J}_{\rm in}}{\dot{M}_{\rm in}},
\end{align}
a general boundary condition on the evolution of CBD can be expressed according to Eq.~\eqref{eq:jdot_in} as 
\begin{align}
    3\pi\nu\Sigma &= \dot{M}_{\rm in}\left(\frac{l_0}{\sqrt{GM_{\rm bin}R_{\rm in}}} -1 \right)\nonumber\\
    &= \dot{M}_{\rm in}\left(\frac{\chi}{\sqrt{a_{\rm bin}/R_{\rm in}}} - 1 \right), \quad r = R_{\rm in} \label{eq:inner_bdry_cond1}
\end{align}
A value $\chi = 0$ corresponds to the well-known no-torque boundary condition, for which no net angular momentum is exchanged between the binary and disk.  For net positive torque on the disk ($\chi < 0$) the radial inflow velocity is reduced below the no-torque value, with the limit $\chi \rightarrow -\infty$ leading to $v_{r}(t, r = R_{\rm in}) = 0$, i.e. a decretion disk \citep{Rafikov16}.

Given $\dot{M}_{\rm d,acc} = -\dot{M}_{\rm bin}$, Eq.~\eqref{eq:adotbin} is integrated to obtain
\begin{align}
    \frac{a_{\rm bin}(t)}{a_{\rm bin, 0}} = \left(1 - \frac{M_{\rm acc}}{M_{\rm d, 0}}\frac{M_{\rm d,0}}{M_{\rm bin,0}}\right)^{2\chi(q+1)^{2}/q - 3}\label{eq:adotbin2},
\end{align}
where $M_{\rm acc}(t) \equiv \int_0^t\,dt^\prime\,\dot{M}(R_{\rm in}) < 0$ is the mass accreted by the binary.  For values of $|\chi|$ of order unity, substantial orbital migration can occur even for modest mass accretion; for example, for $q = 1$, $M_{\rm d,0} \sim 0.2M_{\rm bin,0}$, $\chi = -1$, the binary separation will shrink in half after accreting only $\approx 7\%$ of its mass.

The radial domain on which disk evolves changes in time as a consequence of the exchange of mass and angular momentum with the disk.  Most studies (e.g., \citealt{Gagnier&Pejcha22,Siwek+23}) treat the binary evolution perturbatively, by fixing the binary orbit while evolving the surrounding gas.  Our 1D setup allows us to couple the binary and disk in a more consistent manner. To do so, we make two assumptions: (1) the binary semi-major axis  evolves sufficiently slowly compared to the properties of the disk near its inner edge that $\dot{R}_{\rm in}$ evolves in quasi-static manner and the vertical structure equations hold at all times and radii; (2) the inner boundary of the disk (around which the binary torque is applied) is defined at all times according to
\begin{align}
    \frac{R_{\rm in}(t)}{a_{\rm bin}(t)} = \rm 2.
\end{align}
Under these assumptions, Eq.\eqref{eq:sigma_diff_eq} holds in the time-dependent radial domain $r \in [R_{\rm in}(t), \infty)$ at all times. The radial velocity of the gas with respect to the moving boundary is given by
\begin{align}
    v_{\rm in} = -\frac{R_{\rm in}}{a_{\rm bin}}\dot{a}_{\rm bin} - \frac{3}{\Sigma r^2\Omega}\frac{\partial}{\partial r}\left(\nu\Sigma{r^2\Omega}\right)\bigg\rvert_{\rm r = R_{\rm in}(t)}
\end{align}
Since $\dot{M}_{\rm in} = 2\pi\,R_{\rm in}\Sigma\,v_{\rm in}$, the boundary condition \eqref{eq:inner_bdry_cond1} can be expressed as
\begin{align}
    -
\frac{R_{\rm in}}{a_{\rm bin}}\dot{a}_{\rm bin} &- \frac{3}{\Sigma r^2\Omega}\frac{\partial}{\partial r}\left(\nu\Sigma{r^2\Omega}\right)\nonumber\\
    &= -\frac{3\nu}{2r(1- \chi\sqrt{R_{\rm in}/a_{\rm bin}})}, \quad r = R_{\rm in}. \label{eq:inner_bdry_cond}
\end{align}

\subsection{Initial and Boundary Conditions}
\label{sec:numerical}

We adopt an initial radial density profile of the form
\begin{align}
    \Sigma(t=0, r) = \Sigma_0\,\left(\frac{r}{R_{\rm d,0}}\right)^{-p}\,e^{-r/R_{\rm d,0}} \label{eq:init_data},
\end{align}
which exhibits a power-law form inside a characteristic radius $R_{\rm d,0}$, where $p$, $R_{\rm d,0}$, $\Sigma_0$ are free parameters.  The values of $\Sigma_0$ and $R_{\rm d,0}$ are determined, respectively, by the total initial mass and angular momentum of the disk, which we parameterize as fractions $f_{\rm m}/f_{\rm j}$ of the envelope mass/angular momentum at the onset of CE (Sec.\ref{sec:outline_of_disk_evolution}).  Motivated by \citet{Lau+22a,Lau+22b}, we fiducially assume $f_{\rm m} = 0.1$, $f_{\rm j} = 1$, and $p = 2$, though we explore the effects of varying these choices.  The initial (post-CE) binary properties $\{M_{\rm bin,0}, a_{\rm bin,0}\}$ constitute the final pieces of initial data.

We employ a large radial grid extending to $R_{\rm out} = 10^6\,R_\odot \gg R_g$ (Eq.~\eqref{eq:Rg}) and allow the disk to expand freely into ``vaccuum'' without imposing an outer boundary condition.  In most of our models, the mass contained beyond $r = R_{\rm g}$ is negligible at all times due to high efficiency of photoevaporation mass-loss in truncating the outer disk at radii $\gtrsim R_{\rm g}$ (Sec.~\ref{sec:photoionization}).  

The model parameters are summarized in Table \ref{tab:parameters_symbols}. The full set of equations solved is collated in Appendix \ref{app:summary_equations}, together with the solution method. In summary, we solve Eq.~\eqref{eq:sigma_diff_eq} and Eq.~\eqref{eq:adotbin} for the disk surface density $\Sigma(r,t)$ and binary properties $\{a_{\rm bin}(t),M_{\rm bin}(t)\}$ coupled through the inner boundary condition (Eq.~\eqref{eq:inner_bdry_cond}).

\begin{deluxetable*}{cccc}
\tablecolumns{3}
\tablewidth{0pt}
 \tablecaption{Model Parameters and Representative Symbols \label{tab:parameters_symbols}}
 \tablehead{
 \colhead{Symbol} & \colhead{Parameter} & \colhead{Equation} & \colhead{Fiducial value}}
 \startdata 
  $^{\dagger}M_{\star}$  & Mass of donor star prior to CE & - & $25\,M_\odot$\\
  $^{\dagger}R_{\star}$  & Radius of donor star prior to CE & - & $10^3\,R_\odot$\\
  $^{\dagger}M_{\bullet}$  & Mass of compact accretor (e.g., BH or NS) prior to CE & - & $10\,M_\odot$\\
  $M_{\rm bin,\rm pCE} \equiv M_{\star}+M_{\bullet}$  & Binary total mass prior to CE & - & $35\,M_\odot$\\
  $q_{\rm pCE} \equiv M_{\star}/M_{\bullet}$ & Ratio of donor mass to accretor mass prior to CE & - & 2.5\\
  $a_{\rm RLOF}$ & Binary semi-major axis prior to CE & - & $2170\,R_\odot$\\
  $^{\dagger}a_{\rm bin, 0}$ & Binary semi-major axis at onset of CBD phase (after CE) & Eq.~\eqref{eq:aRLOF} & $30\,R_\odot$\\
  $M_{\rm env} \equiv M_{\star}-M_{\rm c}$  & Donor envelope mass removed during CE & Eq.~\eqref{eq:fm_def} & $15\,M_\odot$\\
   $M_{\rm c}$  & Stripped He core mass (including residual bound envelope) & - & $10\,M_\odot$\\
 $M_{\rm bin,0} \equiv M_{\rm c} + M_{\bullet}$  & Binary total mass at onset of CBD phase (after CE) & - & $20\,M_\odot$\\
 $q \equiv M_{\rm c}/M_{\bullet}$ & Ratio of core mass to accretor mass after CE & - & 1\\
 $\alpha_{\rm CE, 0}\lambda$ & CE efficiency (neglecting CBD interaction) & Eq.~\eqref{eq:alphaCE} & $0.23$\\
 $^{\dagger}\eta_{\rm Edd} \equiv L_{\rm bin}/L_{\rm Edd}$ & Eddington luminosity ratio of post-CE binary & Eq.~\eqref{eq:Lbin} & 0.7\\
 $M_{\rm d, 0}$ & Initial mass of CBD & Eq.~\eqref{eq:fm_def} & $1.5\,M_\odot$\\
 $R_{\rm d, 0} \approx R_{\rm m,1/2}(t= 0)$ & Initial half-mass radius of CBD & Eqs.~\eqref{eq:Rd0}, \eqref{eq:init_data} & $238 \,R_\odot$\\
 $^{\dagger}f_{\rm m} \equiv M_{\rm d,0}/M_{\rm env}$ & Bound envelope mass fraction (that forms the CBD) & Eq.~\eqref{eq:fm_def} & $0.1$\\
 $^{\dagger}f_{\rm j}$ & Ratio of specific angular momentum of CBD to pre-CE binary  & Eq.~\eqref{eq:fj_def} & 1\\
 $^{\dagger}\chi$ & Disk-binary angular momentum coupling efficiency & Eq.~\eqref{eq:acc_eigenvalue} & -1\\
 $^{\dagger}\alpha$ & Disk viscosity parameter & Eq.~\eqref{eq:nu} & $10^{-2}$\\
 $^{\dagger}\xi$ & Advective cooling parameter & Eq.~\eqref{eq:cooling_adv} & 1\\
 $^{\dagger}p$ & Power-law index of initial CBD radial density profile  & Eq.~\eqref{eq:init_data} & 2\\
 \enddata
 \tablecomments{$^{\dagger}$One choice of independent variables which uniquely define the model.}
\end{deluxetable*}

\section{Results}
\label{sec:results}

This section presents our results for the disk/binary evolution.  Sec.~\ref{sec:fiducial_model} describes the fiducial model, as earlier outlined in Sec.~\ref{sec:outline_of_disk_evolution}.  In Sec.~\ref{sec:non_fiducial_models}, we explore the sensitivity of the disk evolution to variations in parameters about the fiducial set.

\subsection{Fiducial Model}
\label{sec:fiducial_model}

\begin{figure*}
    \centering
    \includegraphics[width=1.0\textwidth]{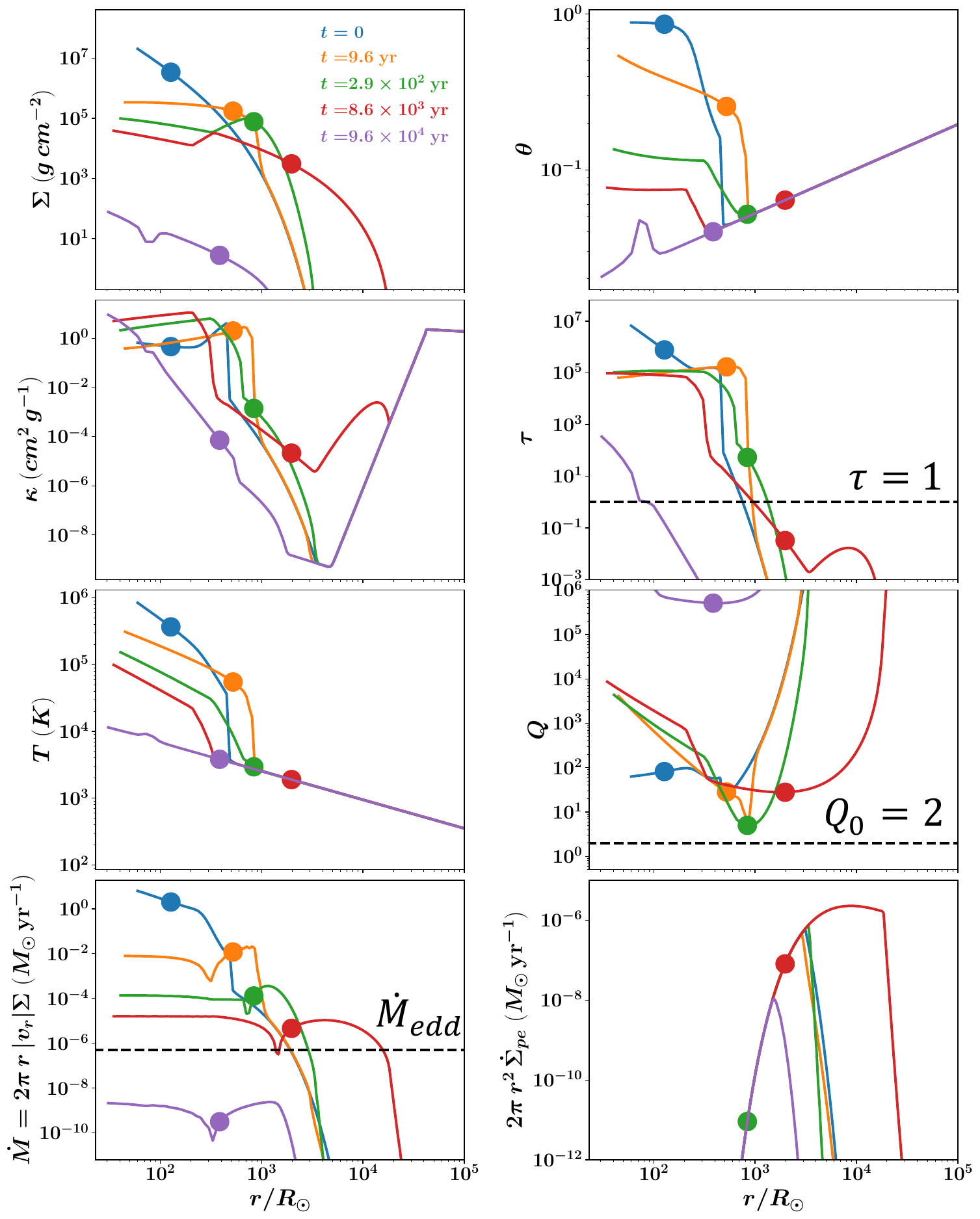}
    \caption{Snapshots of the radial profiles of various properties of the CBD for the fiducial model (see Table~\ref{tab:parameters_symbols}).  The location of the half-mass radius $R_{\rm m,1/2}$ is marked with a dot on each profile.  The expansion of the disk can also be tracked in the plot of mass accretion rate (lower left). Note that the inner boundary of the disk is dynamical: $R_{\rm in}(t=0) = 60\,R_\odot$ and $R_{\rm in}(t\approx 10^5\,\rm yr) \approx 30\,R_\odot$.}
    \label{fig:fiducial_snapshots}
\end{figure*}

The results of the fiducial model are summarized in Fig.~\ref{fig:fiducial_snapshots}.  The model parameters are given in the final column of Table~\ref{tab:parameters_symbols}.  The post-CE binary has a mass-ratio $q = 1$, initial mass $M_{\rm bin,0} = 20M_\odot$, and initial semi-major axis $a_{\rm bin,0} = 30\,R_\odot$.  Taking $R_\star \approx 10^3\,R_\odot$ and $a_{\rm RLOF} \approx 2200\,R_\odot$ for the initial pre-CE donor star, these initial conditions correspond to those achieved by a (partially-failed) CE phase with efficiency parameter $\lambda \alpha_{\rm CE,0} \approx 0.23$ (see Eq.~\eqref{eq:alphaCE}).  
\begin{figure}
    \centering
    \includegraphics[width= 0.48\textwidth]{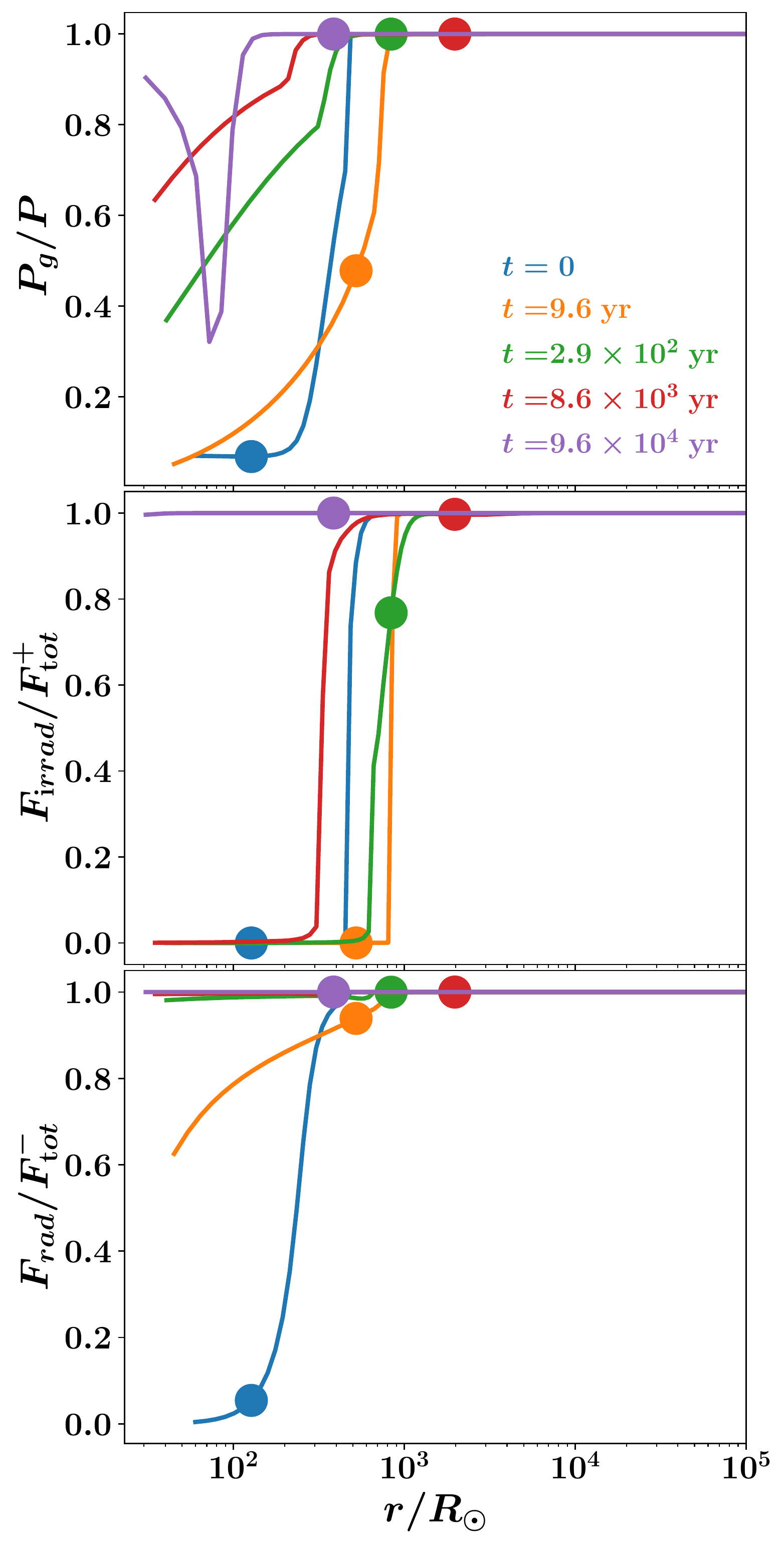}
    \caption{Similar to Fig.~\ref{fig:fiducial_snapshots} but showing for the fiducial model snapshot radial profiles of the ratio of gas pressure to total pressure ($P_{\rm gas}/P$; {\it top panel}); the fraction of the disk heating due to irradiation from the binary ($F_{\rm irrad}/F^+_{\rm total}$, $F^+_{\rm total} \equiv F_{\rm irrad} + F_{\rm visc}$; {\it middle panel}); and the ratio of the radiative cooling rate to the total cooling rate, i.e. the radiative efficiency ($F_{\rm rad}/F_{\rm tot}^{-}$, $F^-_{\rm total} \equiv F_{\rm rad} + F_{\rm adv}$; {\it bottom panel}).  The half-mass radius $R_{\rm m,1/2}$ is marked with a dot on each profile.  The expanding disk evolves from a radiation-dominated, viscously-heated, advectively-cooled state at early times to an irradiation-supported, radiatively-cooled, gas-pressure-supported state at late times.}
    \label{fig:fiducial_disk_fractions_snapshots}
\end{figure}
\begin{figure}
    \centering
    \includegraphics[width=0.48\textwidth]{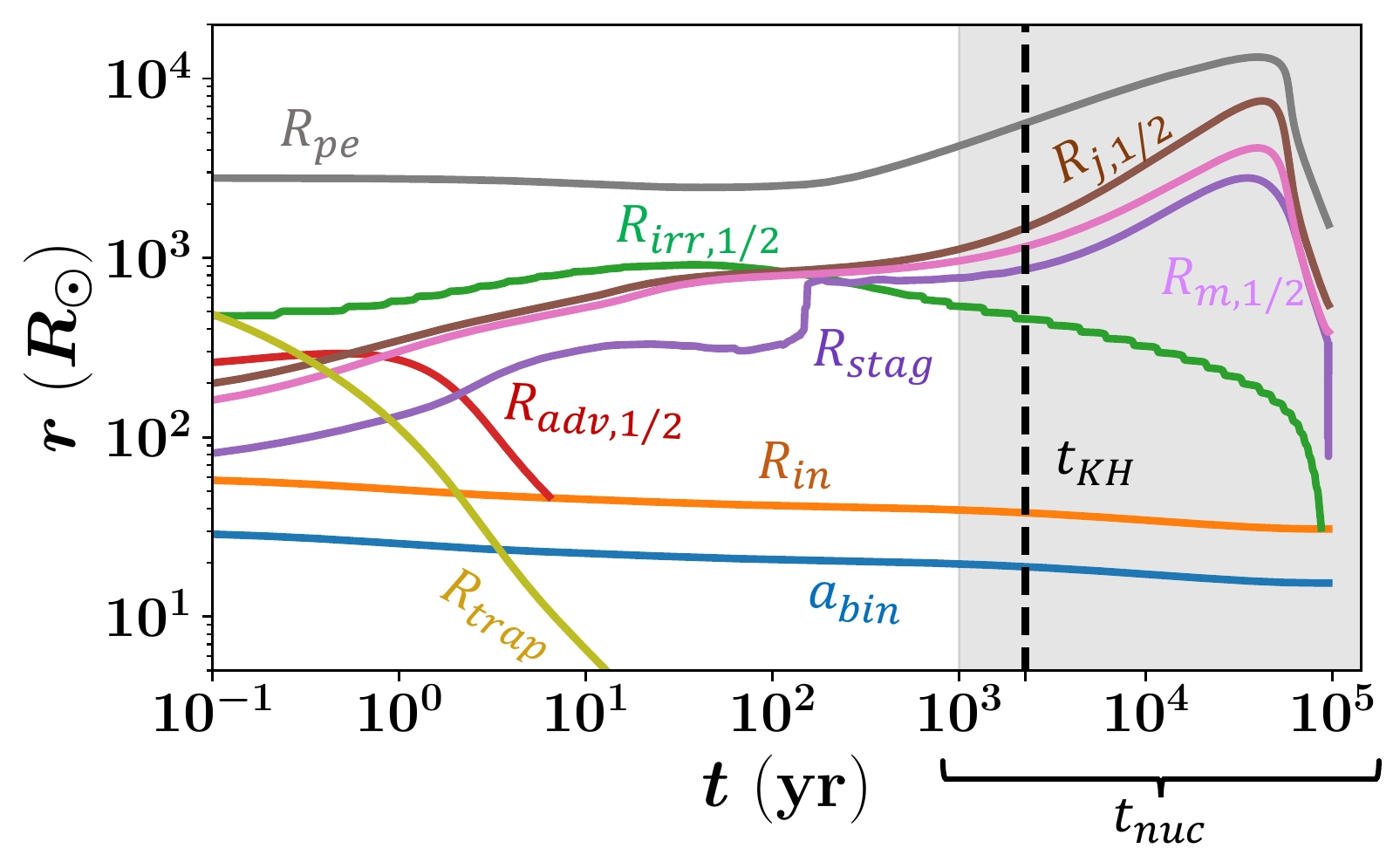}
    \caption{Time evolution of critical radii for the fiducial model.  By assumption of a fixed binary cavity size, the inner edge of the gas disk obeys $R_{\rm in} = 2\,a_{\rm bin}$ at all times.  Inside the trapping radius $R_{\rm trap}$ photons are advected inwards with the accretion flow faster than they can escape via outwards diffusion (Eq.~\eqref{eq:Rtrap}) (olive curve), while within $R_{\rm adv, 1/2}$ (red curve) the rate of advective cooling exceeds half of the total cooling rate.  The stagnation radius $R_{\rm stag}$ is defined as the smallest radius where $v_r = 0$, thus separating gas flowing in towards the binary and that spreading viscously outwards.  The radii $R_{\rm m, 1/2}$ and $R_{\rm j, 1/2}$ are those within which half of the disk mass and angular momentum are contained, respectively. The radius $R_{\rm irr, 1/2}$ is the location beyond which irradiation heating rate exceeds half the total heating rate (i.e. the inner edge of the irradiation-dominated region), while $R_{\rm pe}$ is where the rate of photoevaporation wind mass-loss $2\pi r^{2}\dot{\Sigma}_{\rm pe}$ peaks.  For comparison, we show the thermal time (Eq.~\eqref{eq:t_kh}; vertical dashed line) and an approximate range of nuclear times (gray shaded region) of the stripped He core.}
    \label{fig:fiducial_crit_radii}
\end{figure}

    During its earliest stages of evolution (blue and orange curves in Fig.~\ref{fig:fiducial_snapshots}), the disk can be decomposed into two regions with distinct characteristics. At $r \lesssim 400\,R_\odot$, the disk is hot with $T \gtrsim 10^5\,K$ and geometrically thick with $\theta \gtrsim 0.3$.  In this region radiation pressure dominates over gas pressure.  The disk heating is provided mainly by viscous dissipation, while the disk cools primarily through radial advection rather than radiation ($F_{\rm rad}^{-} \ll F^{-}$) (Fig.~\ref{fig:fiducial_disk_fractions_snapshots}) because of the high optical depth $\tau  \gtrsim 10^5$ and correspondingly long photon diffusion timescale.  At radii $r \lesssim 200\,R_\odot \sim R_{\rm trap}$ (Eq.~\eqref{eq:Rtrap}) advective cooling completely dominates (Fig.~\ref{fig:fiducial_crit_radii}) and the the disk thickness asymptotes to a constant value $\theta \approx 0.9$; in this region, the gas is distributed quasi-spherically and hence height-averaged quantities should be interpreted as spherical averages (see, e.g., \citealt{Narayan&Yi95}).  At $r \approx 400\,R_\odot$, the midplane temperature drops below $T \sim 10^{4}$ K and there is a sharp drop in opacity, which reaches $\kappa < 10^{-8}\,\rm cm^2\,g^{-1}$ by $r \approx 5000\,R_\odot$.  This ``opacity gap'' region of the disk is extremely optically thin $\tau < 10^{-3}$ and the midplane temperature $10^3\,K\lesssim T \lesssim 10^4\,K$ is set by irradiation heating, which follows the expected radial profile $T \propto r^{-3/7}$, independent of $\Sigma$ (see Appendix~\ref{app:analytic}). Radiation is inefficient in providing pressure support such that gas pressure dominates at large radii $r \gtrsim 400\,R_\odot$ (Fig.~\ref{fig:fiducial_disk_fractions_snapshots}).

    Taking as a characteristic disk size the radius $R_{\rm m, 1/2} \approx R_{\rm d,0} \approx 238\,R_\odot$ within which half of the initial disk mass is contained (Fig.~\ref{fig:fiducial_crit_radii}), the corresponding initial viscous time is $t_{\nu,0} \approx 7\,\rm yr$ (Eq.~\eqref{eq:tnu}).  Snapshots of the mass-flux radial profile (bottom lower-left panel of Fig.~\ref{fig:fiducial_snapshots}) demonstrate that the inner disk evolves to a quasi-steady state on this timescale.  An approximately radially-constant mass flux is established at small radii, whose value decays slowly in time.  This steady-state is achieved interior to the stagnation radius (defined as where $v_r = 0$) and half-mass radius, both of which expand in time as angular momentum is transported outwards (Fig.~\ref{fig:fiducial_crit_radii}).  
    
    As the disk loses mass to accretion, radiative cooling quickly becomes efficient across the entire inner disk, which transitions to a geometrically thin state $\theta \ll 1$ by $t \approx 300\,\rm yr$ (Figs.~\ref{fig:fiducial_disk_fractions_snapshots}, \ref{fig:fiducial_crit_radii}), slowing down the disk evolution timescale $t_{\nu} \propto \theta^{-2}$ considerably.  As viscous expansion continues and the disk continues to thin, viscous heating becomes less efficient than irradiation heating across the radii containing most of the disk mass ($F_{\rm irrad} \simeq F^{+}_{\rm total}$; middle-panel of Fig.~\ref{fig:fiducial_disk_fractions_snapshots}), as illustrated also by the dropping value of $R_{\rm irrad,1/2}$ in Fig.~\ref{fig:fiducial_crit_radii} after $t \approx 100$ yr.  Throughout its evolution, the disk is gravitationally stable with $Q > Q_0$ (Eq.~\eqref{eq:Toomre}), as shown in the right-panel of the third row in Fig. \ref{fig:fiducial_snapshots}, though $Q \sim 2$ is achieved across intermediate radii at early times.
    
    The structure and stability of radiation-dominated geometrically-thin accretion disks remain areas of long-standing research (e.g., \citealt{Lightman&Eardley74,Jiang+17}).  At very late times $t\approx 10^5\,\rm yr$ in our fiducial model, radiation pressure temporarily becomes comparable to gas pressure in a narrow region around $r \approx 70\,R_\odot$, where $\Sigma \approx \rm 10\, g\,cm^{-2}$ is achieved (purple curve on the left-panel of Fig.~\ref{fig:fiducial_disk_fractions_snapshots}; see also Fig.~\ref{fig:paramspace}). The disk is marginally optically-thick and irradiation heating dominates under these conditions. However, this transition to a low-density, radiation-pressure-dominated regime is not robust, as we find it depends on uncertain details of the model such as our choice of the interpolation function for the radiation pressure support between the optically-thick and optically-thin limits (Eq.~\eqref{eq:rad_pressure}).  Furthermore, this transitory state is achieved only at late times, after the CBD has already lost almost all of its mass to accretion or winds; hence, conclusions obtained for the earlier evolution of greatest interest are fortunately not affected by these uncertainties.

\begin{figure}
    \centering
    \includegraphics[width=0.48\textwidth]{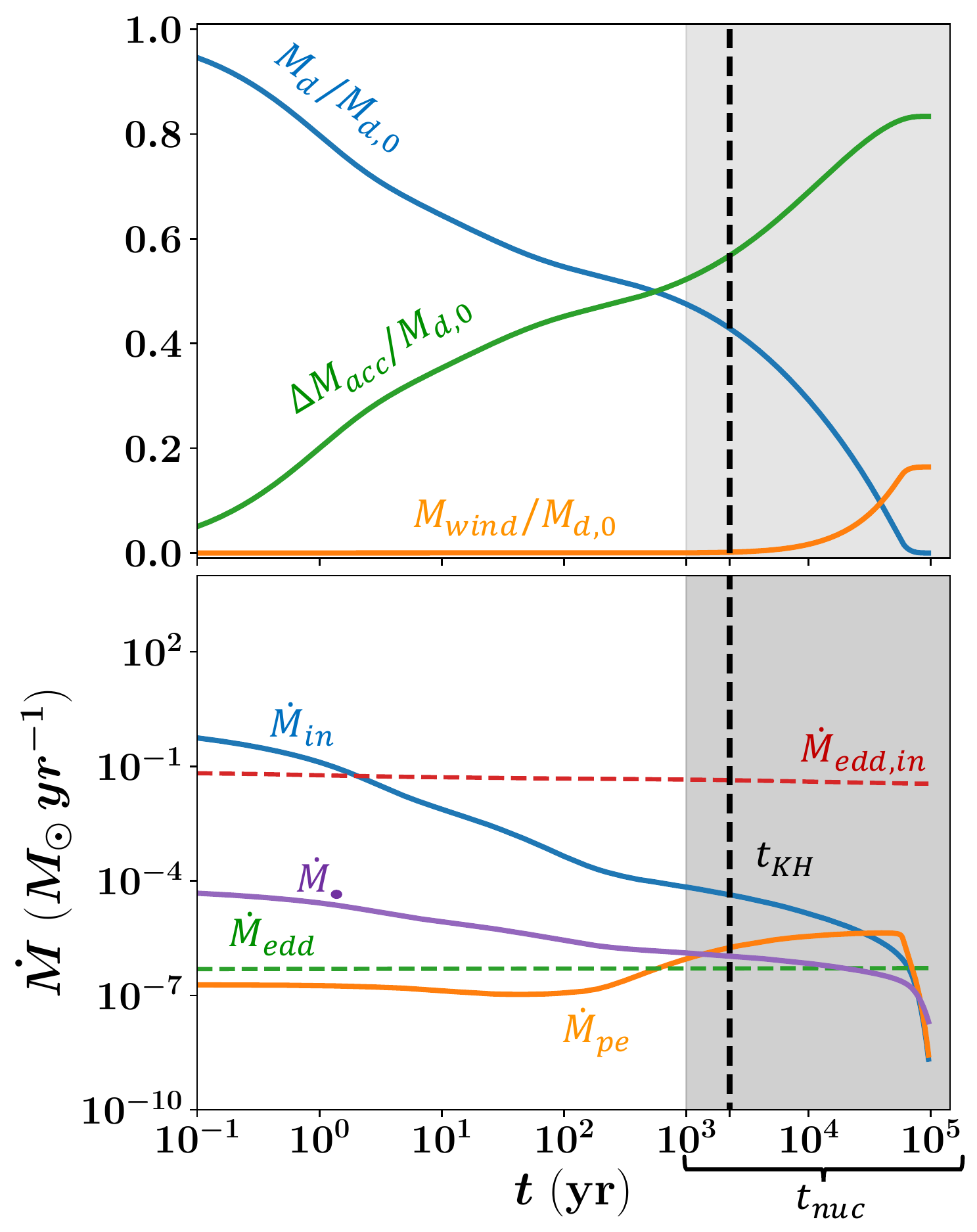}
    \caption{{\it Top panel:} Partition of the initial disk mass and angular momentum among the CBD, binary (through accretion, green curve) and the photoevaporation-wind as a function of time, for the fiducial model.  {\it Bottom panel:} Rates of various mass-loss processes from the CBD, including: $\dot{M}_{\rm in}$, accretion rate onto the binary at $R_{\rm in}$; $\dot{M}_{\bullet}$, accretion rate reaching the central BH/NS at $R_{\rm isco}$ accounting for super-Eddington wind mass-loss (Eq.~\eqref{eq:MdotBH}); $\dot{M}_{\rm pe}$, total mass-loss rate to photoevaporation driven winds (Eq.~\eqref{eq:sigma_dot_pe}).  For comparison we show the Eddington accretion rate $\dot{M}_{\rm edd} \equiv L_{\rm Edd,\bullet}/(0.1c^{2})$ of the BH/NS and the local Eddington accretion rate $\dot{M}_{\rm edd,in} \equiv (L_{\rm Edd}/c^{2})(R_{\rm in}/R_{\rm G})$ at $R_{\rm in}$.  We also show the thermal time (Eq.~\eqref{eq:t_kh}; vertical dashed line) and an approximate range of nuclear times (gray shaded region) of the stripped He core.}
    \label{fig:fiducial_disk_mass_j}
\end{figure}
\begin{figure}
    \centering
    \includegraphics[width=0.48\textwidth]{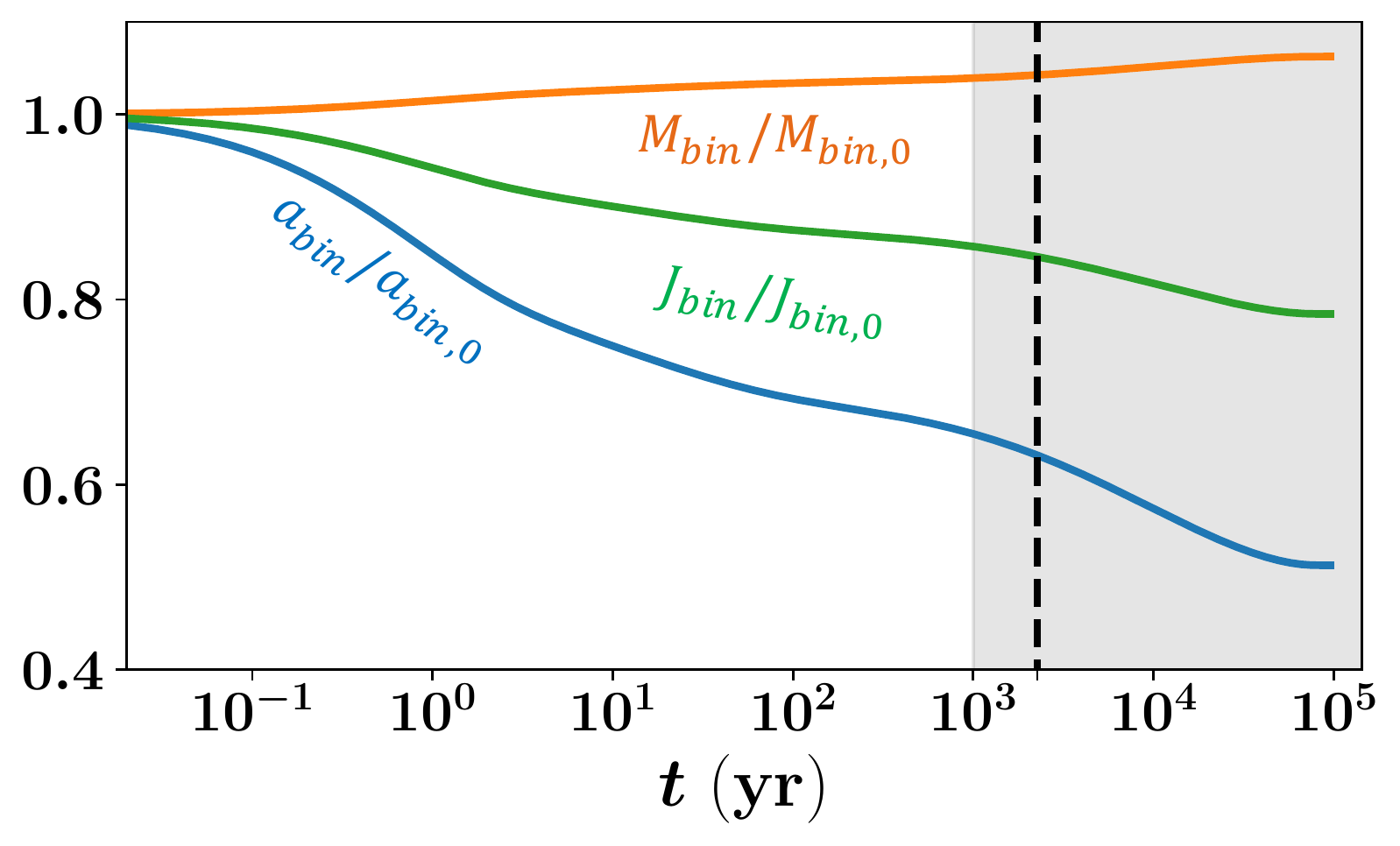}
    \caption{Evolution of the binary total mass $M_{\rm bin}$, orbital angular momentum $J_{\rm bin}$, and semi-major axis $a_{\rm bin}$, for the fiducial model. Note that we assume a circular, equal-mass binary at all times, as described in Sec.~\ref{sec:binaryevo}.  We therefore neglect any mass or angular momentum loss due to winds from the accretion flow onto the NS/BH, though we explore the implications of such mass-loss on the growth and accretion power of the compact object in Sec.~\ref{sec:hypernebulae}.  For comparison, we show the thermal time (Eq.~\eqref{eq:t_kh}; vertical dashed line) and an approximate range of nuclear times (gray shaded region) of the stripped He core.}
    \label{fig:fiducial_binary}
\end{figure}

     The top panel of Fig.~\ref{fig:fiducial_disk_mass_j} illustrates how the initial mass and angular momentum of the disk are partitioned between the disk, binary, and the photoevaporation-driven wind as a function of time.  As the outer edge of the disk viscously expands and begins to approach the radius $R_{\rm g} \approx 6\times 10^{15}$ cm (Eq.~\eqref{eq:Rg}), photoevaporation mass-loss becomes efficient at $t \approx 10^4\,\rm yr$ (the photo-evaporation rate typically peaks at a disk radius $R_{\rm pe} \sim 0.1R_{\rm g} \sim 10^{4}R_{\odot}$; Fig.~\ref{fig:fiducial_crit_radii}).  The total wind mass-loss asymptotes to $\approx 16\%$ of the initial disk mass by $t \approx 10^5\,\rm yr$.  Assuming the disk evolution is not interrupted until reaching this final state, we find that $0.83\,M_{\rm d,0} \approx 1.25\,M_\odot$ is accreted by the binary by this time.  Although the binary masses only grows by $6\%$ through accretion, this leads to $\approx 49\%$ contraction in binary semi-major axis (see Eq.~\eqref{eq:adotbin2} and the surrounding discussion). The evolution of the binary orbital parameters are depicted in Fig.~\ref{fig:fiducial_binary}.

     The bottom panel of Fig.~\ref{fig:fiducial_disk_mass_j} compares the rate of disk mass-loss due to accretion onto the binary at $R_{\rm in}$ ($\dot{M}_{\rm in}$; blue line) to that to winds from the outer radii of the disk ($\dot{M}_{\rm pe}$; orange line).  We also show the mass accretion rate $\dot{M}_{\bullet}$ which reaches the central BH/NS, once wind mass-loss from the super-Eddington inner regions of the disk are taken into account (see Sec.~\ref{sec:hypernebulae} for details).  While $\dot{M}_{\rm in}$ peaks within the first decades of the disk's evolution, the wind mass-loss rate $\dot{M}_{\rm pe}$ only peaks significantly later ($t \sim 10^{4}$ yr) once the outer edge of the disk has had time to viscously spread to the radii $R_{\rm pe} \sim 0.1R_{\rm g} \sim 10^{4}R_{\odot}$ where photoevaporation becomes significant (Fig.~\ref{fig:fiducial_crit_radii}).  While $\dot{M}_{\rm in}$ at all times exceeds the Eddington accretion rate of the assumed BH companion ($\dot{M}_{\rm edd} \equiv L_{\rm Edd,\bullet}/(0.1c^{2})$; green, dotted line), it only briefly exceeds the {\it locally} defined Eddington rate $\dot{M}_{\rm edd,in} \equiv (L_{\rm Edd}/c^{2})(R_{\rm in}/R_{\rm G})$ corresponding to the weaker gravitational binding energy released through accretion rate through radii $\sim R_{\rm in}$ (red, dotted line), where $R_{\rm G} \equiv GM_{\rm bin}/c^{2}$.\footnote{Insofar that $R_{\rm trap}/R_{\rm in} \approx \dot{M}_{\rm edd,in}/\dot{M}_{\rm edd}$, where $R_{\rm trap}$ is the trapping radius (Eq.~\eqref{eq:Rtrap}), the condition $\dot{M}_{\rm in} \sim \dot{M}_{\rm edd,in}$ is equivalent to $R_{\rm in} \sim R_{\rm trap}$ (Fig.~\ref{fig:fiducial_crit_radii}).}

Our implicit assumption that the disk-binary system evolves in isolation may be violated by the expansion of the He core on either its thermal or nuclear evolution timescale (Sec.~\ref{sec:outline_of_disk_evolution}).  We show an estimate of the core thermal time (Eq.~\eqref{eq:t_kh}) with dotted vertical lines in Figs.~\ref{fig:fiducial_disk_mass_j}, \ref{fig:fiducial_binary}.  On this timescale, the CE-stripped core will expand as it re-establishes thermal equilibrium, potentially leading to RLOF onto the BH/NS companion and the possibility of unstable mass-transfer and a dynamical merger (Sec.~\ref{sec:stability}).  Even if a merger is avoided, the core will continue to evolve towards core-collapse on its nuclear timescale, which is typically $t_{\rm nuc} \sim 10^3 - 10^{5} \,\rm yr$, depending on the original evolutionary state of the donor at the onset of the CE.  Either a binary merger or relatively prompt core-collapse explosion could obviously truncate the binary evolution prior to the final stage we simulate.  For our fiducial model, by $t\approx t_{\rm KH}$ the binary orbit has shrunk by $\approx 37\%$, while the wind mass-loss up to this point is negligible (photoevaporation of the disk only becomes significant after $t \approx 10^4\,\rm yr$).

\subsection{Variations About the Fiducial Model}\label{sec:non_fiducial_models}
 
    The model presented in Sec.~\ref{sec:fiducial_model} has 10 free parameters (one combination of which is denoted by $\dagger$ in Table~\ref{tab:parameters_symbols}).  Here we investigate the sensitivity of models fixing $M_{\rm bin,0} = 20\,M_\odot$ and $q_{\rm bin} = 1$ to variations in the remaining $8$ parameters.  We find that the latter can roughly be divided into two types: (1) those parameters that affect the final state directly, even absent externally-imposed constraints on the disk lifetime; and (2) those that only affect the final state indirectly by controlling the speed of the disk's evolution and hence the state it achieves given some fixed lifetime set by the core's thermal or nuclear evolution. 

\begin{figure*}
    \centering
    \includegraphics[width=1.0\textwidth]{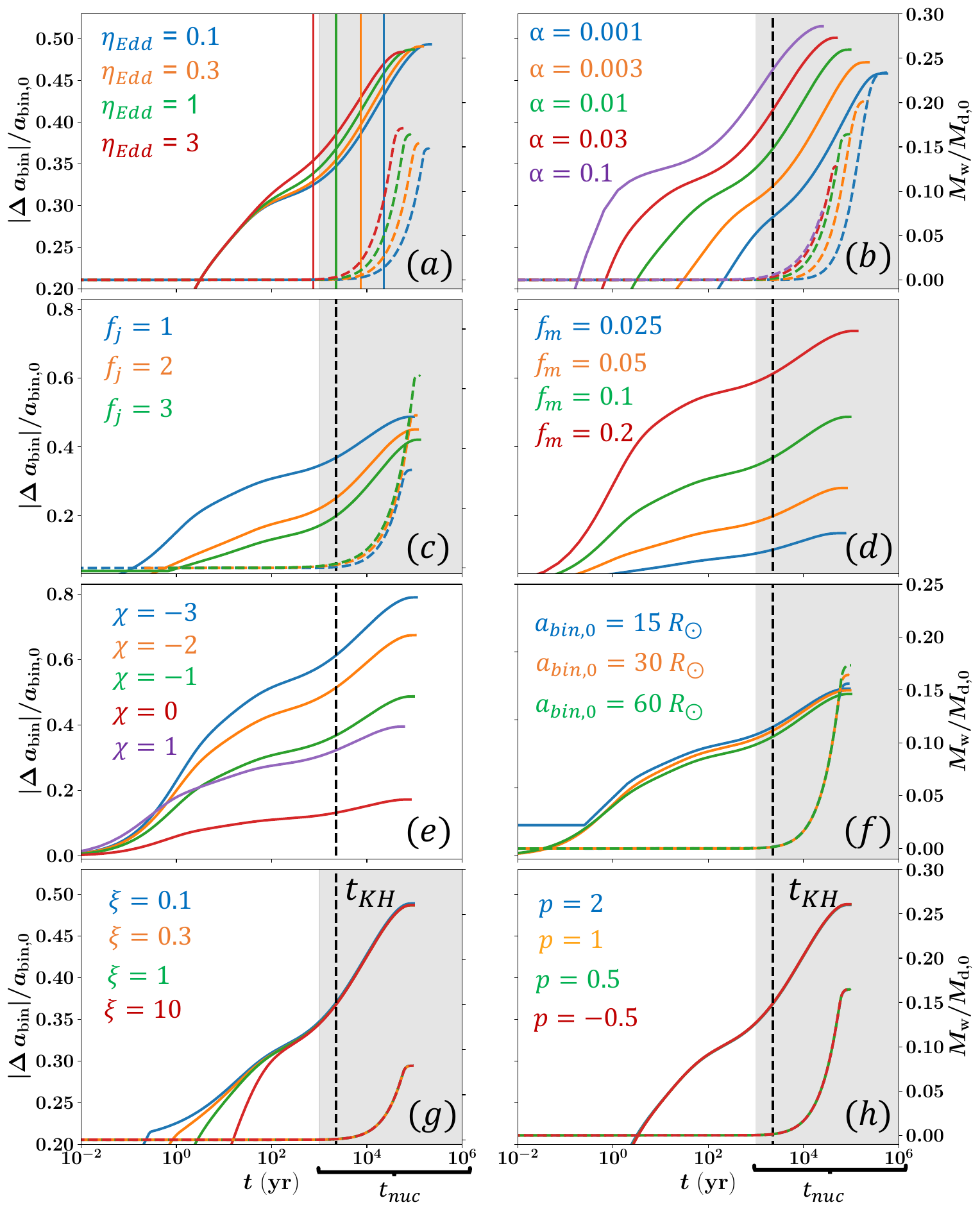}
    \caption{Effect of varying parameters of the model about their fiducial values on the cumulative change of the binary separation $\Delta a_{\rm bin}$ (solid lines; left axes) and cumulative wind mass-loss $M_{\rm w}$ (dashed lines; right axes).  For comparison, we show the thermal time (Eq.~\eqref{eq:t_kh}; vertical dashed line) and an approximate range of nuclear times (gray shaded region) of the core.}
    \label{fig:non_fiducial}
\end{figure*}
    Fig.~\ref{fig:non_fiducial} shows the time evolution of the fractional change in the binary semi-major axis (left vertical-axis, solid curves) as well as the cumulative wind mass-loss normalized to the initial disk mass (right vertical-axis, dotted curves); each panel shows the result of separately varying different parameters about their fiducial model values (Table~\ref{tab:parameters_symbols}).  The estimated thermal timescale and range of nuclear timescales of the stripped core, over which the disk evolution may be prematurely truncated by a merger or core-collapse event, are again shown for comparison with vertical dashed lines and solid shaded regions, respectively.

    Panel (a) of Fig.~\ref{fig:non_fiducial} shows the effect of changing the binary luminosity incident on the disk $0.1L_{\rm edd} < L_{\rm bin} < 3L_{\rm edd}$.  This variation accounts for uncertainties associated with the evolutionary state and luminosity of the core as well as the radiative efficiency and geometry (e.g., effects of self-shielding) of the (super-Eddington; bottom panel of Fig.~\ref{fig:fiducial_disk_mass_j}) BH/NS accretion disk.  By the end of the mass-loss evolution ($t\approx 10^5\,\rm yr$), the fractional binary contraction for each model matches within $\pm 1\%$ of the fiducial model value $\approx 49\%$. The fractional binary contraction at time $t \sim t_{\rm KH}$ (where $t_{\rm KH} \propto L_{\rm bin}^{-1}$ is shown with vertical dotted lines Fig.~\ref{fig:fiducial_disk_mass_j}) exhibits a larger variation $35\% - 43\%$, with a lower binary luminosity leading to more orbital contraction, simply because $t_{\rm KH}$ correspond to a later stage in the disk evolution, at the end of which more of the disk mass is accreted. By contrast, the fractional binary contraction at any fixed time $10^2 \lesssim t \lesssim 10^4$, is smaller for less luminous binaries. This behavior results because disks which experience less irradiation-heating are thinner in regions where irradiation heating dominates over viscous heating, leading to longer viscous timescales and slower disk evolution.  The disk spreads to become irradiation-dominated across most of its mass by $t \approx 100\,\rm yr$ ($R_{\rm m, 1/2} \approx R_{\rm irrad, 1/2}$ in Fig.~\ref{fig:fiducial_crit_radii}), which is indeed roughly the time at which the binary contraction rates shown in Fig.~\ref{fig:non_fiducial} begin to diverge from one another.
    
    The greater wind mass-loss rate associated with a higher luminosity (Eq.~\eqref{eq:sigma_dot_pe}) might naively be expected to lead to greater total wind mass-loss, leading to less accreted mass and concomitant binary contraction.  However, a greater central luminosity also leads to faster disk evolution for reasons argued above, thereby reducing the time the disk spends near the radii $r \approx 0.1 R_{\rm g}$ where the photoevaporation rate peaks.  This competition apparently results in the total wind mass-loss increasing only weakly with binary luminosity (from $0.15\,M_{\rm d,0}$ for $L_{\rm bin}/L_{\rm edd} = 0.1$ to $0.17\,M_{\rm d,0}$ for $L/L_{\rm edd} = 3$), such that the final binary separation is largely independent of $L_{\rm bin}$.

    The role of the viscosity parameter $\alpha$ in determining the speed of disk evolution is more pronounced. Because the viscous time obeys $t_\nu \propto \alpha^{-1}$, smaller values of $\alpha$ lead to slower disk evolution (panel $(b)$ in Fig.~\ref{fig:non_fiducial}). Hence, at a fixed time, disks with lower $\alpha$ result in less binary contraction than for larger $\alpha$; for example, the fractional binary contraction by $t \sim  t_{\rm KH}$ is $\approx 29\%$ (corresponding to an accreted mass $\Delta\,M_{\rm acc} \approx 0.42\,M_{\rm d,0}$) for $\alpha = 0.001$, compared to $\approx 46\%$ ($\Delta\,M_{\rm acc} \approx 0.77\,M_{\rm d,0}$) for $\alpha = 0.1$.  Disks with lower $\alpha$ also lose a greater fraction of their initial mass to photoevaporation-driven wind relative to accretion, a trend which results because photoevaporation has longer to operate when the disk evolution is slower.  However, this does not lead to significant changes in the final state of the binary: for $\alpha= 0.001$ the disk loses only $\approx 10\%$ more of its initial mass to photoevaporation than for $\alpha= 0.1$, resulting in the final binary separation differing by $\approx 5\%$ of $a_{\rm bin,0}$.

    Panel $(c)$ of Fig.~\ref{fig:non_fiducial} explores the dependence on the initial angular momentum of the CBD, as parametrized through the dimensionless parameter $f_{\rm j}$ (Eq.~\eqref{eq:fj_def}). Because the characteristic initial disk radius obeys $R_{\rm d,0}\propto f_{\rm j}^2$ (Eq.~\eqref{eq:Rd0}), larger values of $f_{\rm j}$ (larger initial disk sizes) enhance the role of photoevaporation mass-loss by reducing the time the disk requires to viscously expand to the radii $r \approx 0.1 R_{\rm g} \sim 10^{4}R_{\odot}$ where photoevaporation peaks ($R_{\rm pe}$ in Fig.~\ref{fig:fiducial_crit_radii}). For $f_{\rm j} = 1$ (fiducial model), $R_{\rm d,0} \approx 238\,R_\odot$, and majority of the disk's mass is accreted before its outer edge reaches $10^4\,R_\odot$, leading to negligible wind mass-loss by $t \approx t_{\rm KH}$ and the binary semi-major axis contracting by $37\%$; wind mass-loss starts rising at $t \approx 10^4\,\rm yr$, and approaches $0.16\,M_{\rm d,0}$ at late times $t\approx 10^5\,\rm yr$. On the other hand, for $f_{\rm j} = 3$ ($R_{\rm d,0} \approx 1.2\times10^4\,R_\odot$), disk winds become efficient earlier, reaching $M_{\rm w} \approx 0.32\,M_{\rm d,0}$ and $a_{\rm bin} \approx 0.42\,a_{\rm bin, 0}$ by $t\approx 10^5\,\rm yr$.
    
    The dominant roles of the initial disk mass (as parameterized by $f_{\rm m}$; Eq.~\eqref{eq:fm_def}) and  torque efficiency $\chi$ (Eq.~\eqref{eq:acc_eigenvalue}) in the system evolution are illustrated in panels $(d)$ and $(e)$ of Fig.~\ref{fig:non_fiducial}.  For $\chi = -1$, the orbital separation achieved by $t \approx 10^5\,\rm yr$ spans a wide range of values $0.15\,a_{\rm bin,0} \lesssim a_{\rm bin} \lesssim 0.73\,a_{\rm bin,0}$ for initial disk masses $0.35\,M_\odot \lesssim M_{\rm d,0} \lesssim 3\,M_\odot$.  In these models, we encountered gravitational instability only for the $M_{\rm d,0} = 3\,M_\odot$ ($f_{\rm m} = 0.2$) model for a brief phase of its early evolution (during which the gravitoturbulence condition $t_{\rm cool}\Omega > 1$ described in Sec. \ref{sec:grav_instability} was satisfied).  For the $M_{\rm d,0} = 1.5\,M_\odot$ model, the orbit shrinks by $\approx 79\%$ for $\chi = -3$, compared to a $\approx 17\%$ expansion for $\chi = 1$.  
    More massive disks exhibit slightly longer lifetime and lose a smaller fraction of their initial mass to winds.  
    
    The parameter $\chi$ controls the mass the binary must accrete for the disk to expand to a given radius.  The total angular momentum and mass of a Keplerian disk are related according to $J_d \propto M_{d}\,R_{d}^{1/2}$, where $R_d$ is the characteristic disk radius (e.g., $R_{j, 1/2}$ shown in Fig.~\ref{fig:fiducial_crit_radii}).  Because less angular momentum remains in the disk for larger $\chi$, the disk mass $M_d \propto J_{d}R_{d}^{-1/2}$ remaining after the disk has spread to a fixed radius also decreases with increasing $\chi$.  Panel $(e)$ of Fig.~\ref{fig:non_fiducial} shows that while the wind mass-loss begins to rise simultaneously for all models, the value it saturates at is smaller for larger $\chi$.  This is because for larger $\chi$ less mass remains to be photoevaporated by the time the disk expands to radii $\sim R_{pe} \approx 0.1R_{\rm g}$ at which wind mass-loss sets in.  Given the importance of $f_{\rm m}$ and $\chi$ on the system evolution, we vary them simultaneously in Fig.~\ref{fig:delta_alpha_CE} to explore the range of possible outcomes and their implications for the final post-CE separation (see Sec.~\ref{sec:alpha}).
    
    The dependence of disk/binary evolution on the initial binary separation $a_{\rm bin,0}$ as set by the prior CE phase is investigated in panel $(f)$ of Fig.~\ref{fig:non_fiducial}. Insofar that the disk-driven change in binary separation is expressed as a fraction relative to $a_{\rm bin,0}$, the results agree with each other to within few percent at all times for all values of $a_{\rm bin,0}$; this is likewise the case for the cumulative wind mass-loss.
    
    Panel $(g)$ of Fig.~\ref{fig:non_fiducial} shows that the majority of the disk's evolution is robust to uncertainties associated with our simplified treatment of advective cooling (encapsulated by the parameter $\xi$; Eq.~\eqref{eq:cooling_adv}).  Our fiducial model predicts a wide range of values $\xi = |d{\rm ln} s/d{\rm ln} r|$ between $0.1$ and $10$ are achieved at times $t \lesssim 10\,\rm yr$ and radii $r \lesssim 60\,R_\odot$ where advective cooling dominates the disk thermal balance.  However, insofar that advective cooling only dominates the evolution at early times $t \lesssim 10^2\, \rm yr$ (Fig.~\ref{fig:non_fiducial}, panel $(g)$), the final values for the accreted mass and binary contraction by $t \approx 10^2\,\rm yr$ are largely insensitive to the chosen value $\xi$.  Thus, despite many uncertainties associated with the early advective, locally super-Eddington accretion phase, our results at later times $\gtrsim 10^{3}$ yr (e.g., over which the core evolves), are not sensitive to the details of how this early phase is treated. Lastly, we varied the power-law index $p$ of the density profile (Eq.~\eqref{eq:init_data}) in panel $(h)$ of Fig.~\ref{fig:non_fiducial}, again finding nearly identical results, provided that $R_{\rm d,0}$ and $\Sigma_0$ are determined at fixed $f_{\rm m}$ and $f_{\rm j}$. Our results are therefore also insensitive to the detailed radial mass distribution of the post-CE state, at least for initial density profiles broadly similar to Eq.~\eqref{eq:init_data}.

\section{Applications}
\label{sec:discussion}

\subsection{Implications for the Final Binary Orbit}
\label{sec:alpha}

The efficiency of the CE phase in tightening a binary is typically characterized by the dimensionless parameter \citep{vandenHeuvel76, Tutukov&Yungelson79, Iben&Tutukov84, Webbink84, Livio&Soker88}
\be
\alpha_{\rm CE} \equiv \frac{E_{\rm bind}}{\Delta E_{\rm orb}},
\label{eq:alphaCE}
\ee
where 
\be
\Delta E_{\rm orb} = \frac{GM_{\star}M_{\bullet}}{2a_{\rm RLOF}} - \frac{GM_{\rm c}M_{\bullet}}{2a_{\rm bin,f}}
\ee
is the change in orbital energy from the initial state at RLOF to the final binary separation $a_{\rm bin,f}$ (the {\it initial} separation $a_{\rm bin,0}$ for our CBD calculations).  The envelope binding energy $E_{\rm bind}$ is frequently parameterized according to \citep{deKool87}
\be
E_{\rm bind} = -\frac{GM_{\rm env}M_{\star}}{\lambda R_{\star}} = -\frac{G(M_{\star}-M_{\rm c})M_{\star}}{\lambda R_{\star}}.
\ee
A typical value $\lambda = 0.5$ is assumed in population synthesis modeling  (e.g., \citealt{Hurley+02}), though a wider range $\lambda \sim 0.1-1$ is motivated by 1D stellar structure models for different stellar masses and evolutionary states (e.g., \citealt{Xu&Li10, Loveridge+11, Claeys+14, Kruckow+16, Renzo+23}).  A given set of model parameters $\{ M_{\star},R_{\star},M_{\rm c},M_{\bullet},f_{\rm m}\}$ thus defines a direct mapping $\lambda \alpha_{\rm CE} \leftrightarrow a_{\rm bin,f}.$  

\begin{figure}
    \centering  \includegraphics[width=0.48\textwidth]{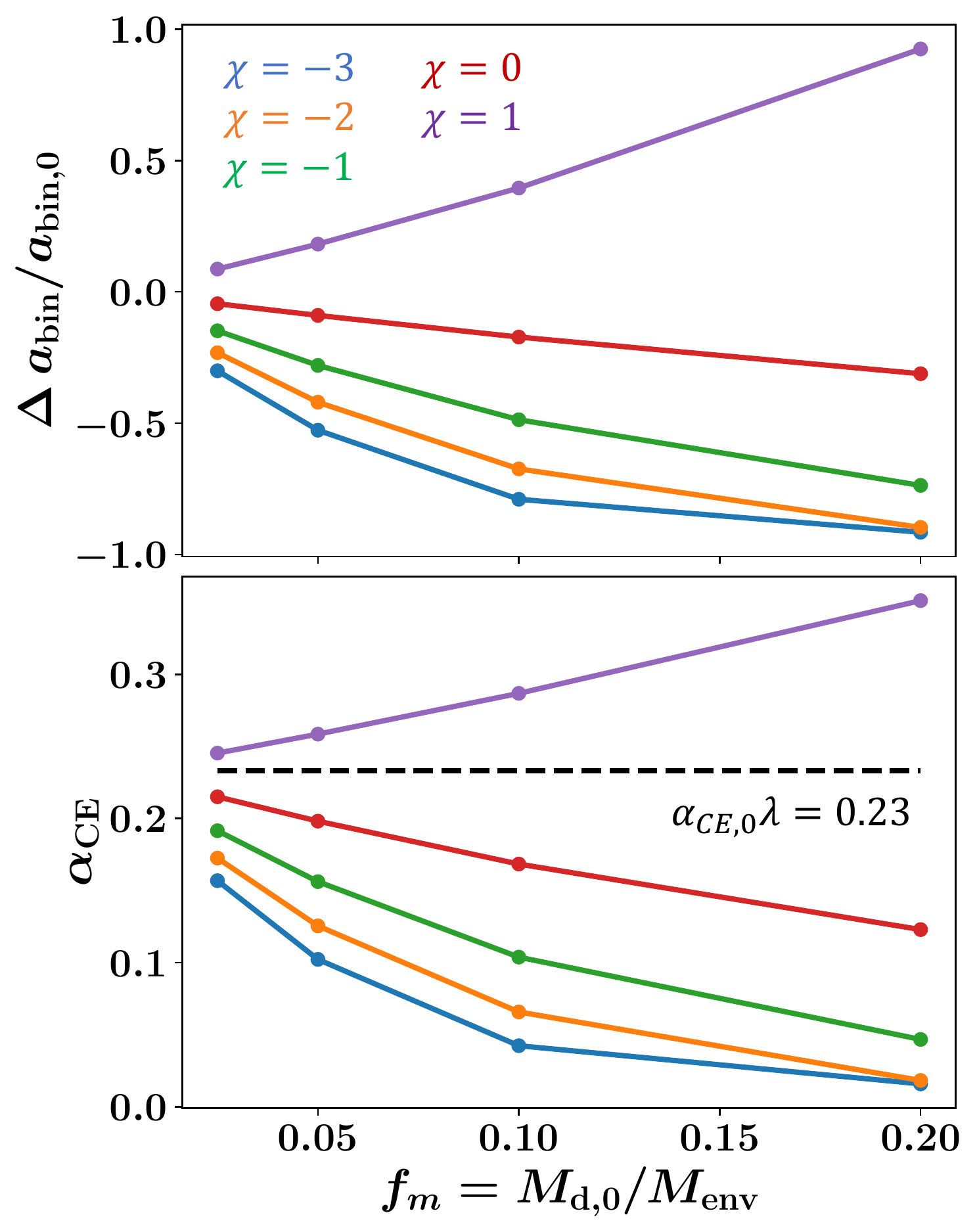}
    \caption{{\it Top panel:} Fractional change in binary semi-major axis $a_{\rm bin,0}$ at $t\approx 10^5\,\rm yr$ due to accretion-driven binary migration during the CBD phase. Results are shown for the fiducial model (Table.~\ref{tab:parameters_symbols}), varying $f_{\rm m}$ (hence the initial disk mass $M_{\rm d,0}$) and binary-CBD torque coupling efficiency $\chi$.  We terminate our simulations when $a_{\rm bin} = 2.5\,R_\odot$ is achieved, which is roughly the RLOF radius for a He core of radius $R_c \approx R_\odot$ and $q = 1$. The models with $f_{\rm m} = 1$ and $\chi = -3$ and $\chi = -2$ reach this final orbital separation by $t\approx 10^5$ {\it Bottom panel:} Corresponding change in the effective value of the CE efficiency parameter $\lambda \alpha_{\rm CE}$ (Eq.~\ref{eq:alphaCE}) compared to the value $\lambda \alpha_{\rm CE,0}$ at the immediate post-CE phase (black, dotted line)}
    \label{fig:delta_alpha_CE}
\end{figure}

The bottom panel of Figure \ref{fig:delta_alpha_CE} shows the change in the {\it effective} value of $\lambda \alpha_{\rm CE}$ which results from the post-CE disk phase, as calculated assuming fiducial model parameters but for a range of initial disk mass-fractions, $f_{\rm m}$.  As expected, greater initial disk masses lead to greater post-CE migration and a larger change in $\lambda \alpha_{\rm CE}$, reducing it to $\lambda \alpha_{\rm CE} \approx 0.04$ from its fiducial value $0.23$ for $f_{\rm m} \gtrsim 0.2$ and $\chi = - 1$. We may conclude that if the CE leaves a non-negligible gaseous disk, the values of $\lambda \alpha_{\rm CE}$ extracted from simulations which follow just the early dynamical phase may differ considerably from their ``true" asymptotic values.  

As already discussed in Sec.~\ref{sec:introduction}, there are observational indications that the standard $\alpha_{\rm CE}$ formalism provides an incomplete description of the final outcome of CE interaction (e.g., \citealt{Roepke&DeMarco22,Hirai+Mandel22,Wilson&Nordhaus22}).  As an example in the massive-star case, reproducing the compact object merger rate observed by LIGO through the isolated binary evolution channel appears to require value of $\alpha_{\rm CE} \gtrsim 2$ (e.g., \citealt{Garcia+21,Zevin+21,Broekgaarden&Berger21}), potentially implicating additional source(s) of energy to unbind the envelope than just the gravitational potential released by the shrinking binary.    

However, our results show that if the net binary disk torque is positive ($\chi > \chi_{\rm crit}$), long-term disk-binary interaction can expand the binary orbit (Fig.~\ref{fig:delta_alpha_CE}), populating otherwise ``forbidden'' regions of orbital period according to the standard $\alpha_{\rm CE} < 1$ formalism.  In the low-mass case, such a scenario is consistent with the observation that the widest post-AGB binary systems possess the most massive long-lived disks \citep{vanWinckel18}, preferentially detectable to later times through their infrared emission (e.g., \citealt{Bujarrabal+18}).

Accretion from a CBD can also impact the binary eccentricity (e.g., \citealt{Dermine+13,Izzard&Jermyn18,Gagnier&Pejcha22}).  Moderate eccentricities $e \simeq 0.15$ are observed in a fraction post-CE binaries \citep{Delfosse+99,Edelmann+05,Kawka+15}, extending to $e \simeq 0.5$ in long-period post-AGB systems (see \citealt{vanWinckel18}; their Fig.~1.2). 
Although our calculations assume circular orbits for simplicity, the disk evolution is largely dominated by the viscous evolution occurring at radii $\gg R_{\rm in}$ and hence would not be qualitatively sensitive to the inclusion of a moderate eccentricity, except through the eccentricity-dependence of the binary torque $\chi$.  Simulations of CBDs show that binary torques driven eccentricity growth \citep{Artymowicz+91} towards an equilibrium value $e \sim 0.1-0.5$ that depends on the binary mass ratio (e.g., \citealt{Zrake+21,Siwek+23}).  Insofar that the binary need only accrete $\sim 10\%$ of its mass to reach such an equilibrium eccentricity (e.g., \citealt{Siwek+23}), we should expect significant eccentricities in the massive star case.  Disk-excited eccentricity would operate in addition to any eccentricity of primordial origin \citep{Glanz&Perets21} or that excited by fast episodic mass-loss during late stages of the CE phase \citep{Soker00,Clayton+17,Kashi&Soker18}.  

\subsection{Stability of Post-CE Binary}
\label{sec:stability}

For sufficiently tight post-CE binary separations, the core will undergo RLOF onto the compact companion after expanding on its thermal or nuclear timescale (so-called case BB or case AB mass-transfer, depending on the residual hydrogen content of the core; e.g., \citealt{Delgado&Thomas81,Dewi+02,Ivanova+03,Tauris+17,VignaGomez+22}).  The timing of this second round of RLOF can be accelerated if the CBD causes binary shrinkage, as studied here. 

After RLOF commences the inner binary properties will evolve under the action of mass and angular momentum transfer between the binary components, in addition to the external disk torques we have modeled.  Of particular relevance to the final fate of the system is the stability of this mass-transfer process (e.g., \citealt{Dewi&Pols03,Ivanova+03,Romero-Shaw+20}).  \citet{Tauris+15} find case BB mass-transfer is usually stable, though the conditions for stability under case AB mass-transfer are less clear \citep{Ivanova11,Quast+19,VignaGomez+22}.  However, most if not all of these past studies neglect the effects a relic CBD from an earlier CE would have on the mass-transfer process.  Our results show that this assumption is questionable because the disk lifetime can in general exceed the thermal timescale, if not the nuclear timescale, of the stripped core (e.g., Fig.~\ref{fig:fiducial_disk_mass_j}).

The potential effects of a CBD on the mass-transfer process are at least three-fold.  Firstly, insofar that the accretion rate $\dot{M}_{\rm in}$ onto the binary is significant compared to the donor's mass-loss rate through the $L_1$ Lagrange point, shock interaction between the two accretion streams may occur near the outer edge of the mini-disk feeding the BH or NS.  While it is not possible to assess the impact of such interaction without costly hydrodynamic simulations, we speculate that this process would reduce the efficiency with which angular momentum is transported from BH/NS mini-disk back into the binary orbit through tides; such non-conservative behavior would act to destabilize the mass-transfer process compared to otherwise equivalent CBD-free systems.  A second potential effect of a CBD is the existence of significant binary eccentricity (Sec.~\ref{sec:alpha}), which can also affect the stability of the mass-transfer process (e.g., \citealt{Dosopoulou&Kalogera16a,Dosopoulou&Kalogera16b}).

Finally, the loss of angular momentum from the binary to the CBD can accelerate the mass-transfer rate from the He core to the compact companion relative to an otherwise equivalent isolated binary.  Such an increase in the donor mass-loss rate will also tend to destabilize the binary insofar that (1) the donor's surface layers are more prone to expand adiabatically upon mass-loss than to stay in thermal equilibrium (e.g., \citealt{Ge+20}); (2) mass-loss can take place through the $L_2$ Lagrange point as well as from $L_1$ (e.g., \citealt{Pejcha+16b,Pejcha+16a,Linial&Sari17,Lu+23}), increasing the specific angular momentum loss (though the interaction of $L_2$ mass-loss stream with inflowing material from the outer disk might suppress this process).
\begin{figure}
    \centering
    \includegraphics[width= 0.48\textwidth]{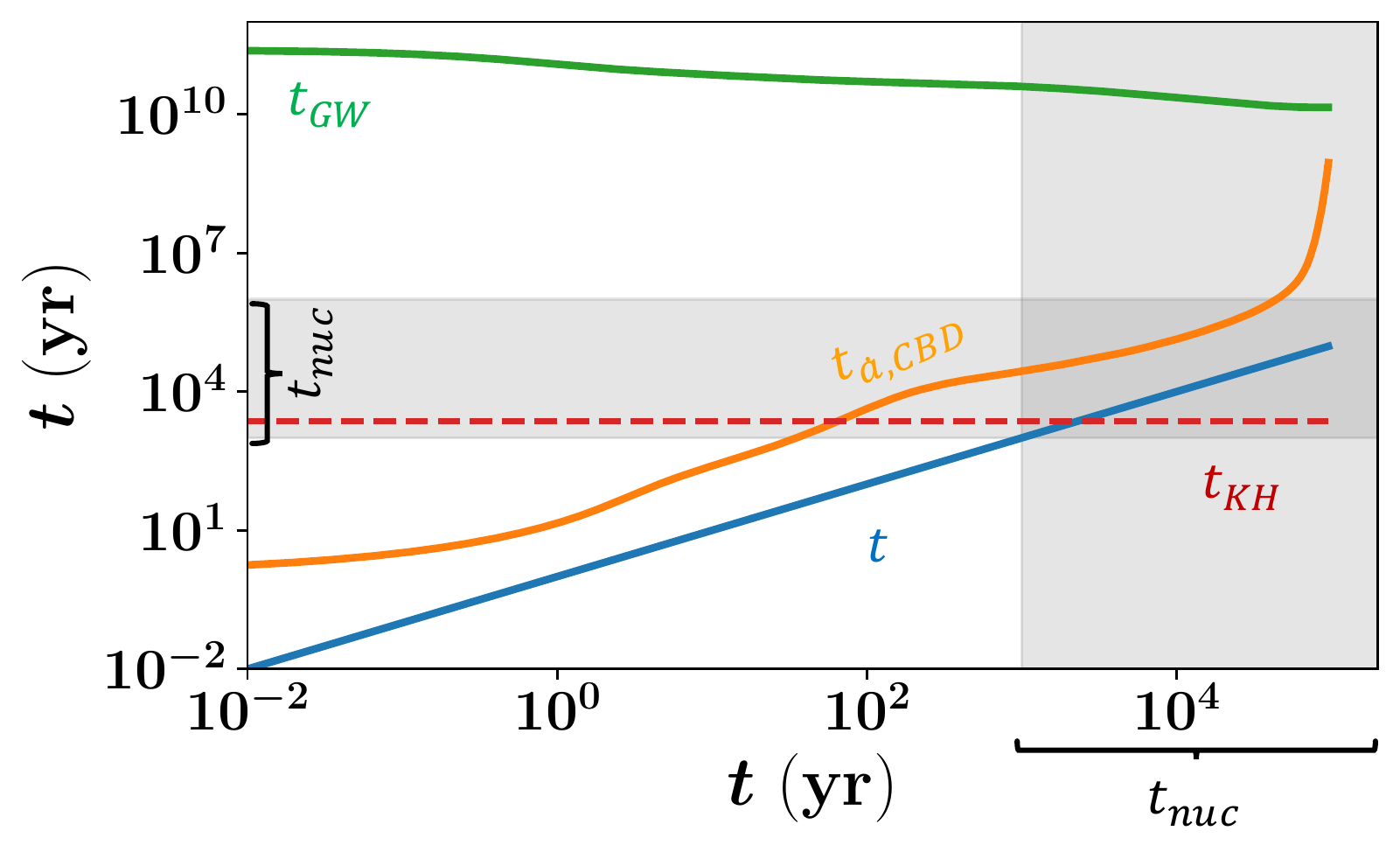}
    \caption{Comparison of timescales relevant to the mass-transfer evolution of the post-CE binary as a function of elapsed time since CBD formation (blue line) for our fiducial model.  A green line shows the timescale for gravitational wave-driven inspiral, while $t_{\dot{a},\rm CBD} \equiv a_{\rm bin}/\dot{a}_{\rm bin}$ shows the rate of binary shrinkage due to interaction with the CBD.  If the He core begins mass-transfer onto the BH/NS companion at early times when $t_{\dot{a}, \rm CBD} < t_{\rm KH}, t_{\rm nuc}$, angular momentum loss to the CBD could substantially accelerate the mass-transfer rate of the binary relative to the CBD-free case, potentially affecting its dynamical stability and leading to tidal disruption of the He core.}
    \label{fig:fiducial_timescales}
\end{figure}

Fig.~\ref{fig:fiducial_timescales} compares for the fiducial model various timescales of relevance to binary evolution as a function of time after the CE.  If the core only begins RLOF as a result of envelope expansion on the thermal timescale $t_{\rm KH}$ (Eq.~\eqref{eq:t_kh}), then at such an epoch (where the blue and red lines cross in Fig.~\ref{fig:fiducial_timescales}) the timescale for disk-driven binary contraction $t_{\dot{a},\rm CBD} \equiv a_{\rm bin}/\dot{a}_{\rm bin}$ is generally comparable or greater than $t_{\rm KH}$.  In such cases, the mass-transfer rate would presumably be altered only moderately compared to the otherwise equivalent disk-free case (for which mass-loss is driven by thermal-expansion of the core, at an approximate rate $\dot{M} \sim M_{\rm c}/t_{\rm KH}$).  On the other hand, if RLOF begins on a timescale much shorter than the core's thermal timescale ($t_{\dot{a},\rm CBD} \ll t_{\rm KH}$), the resulting mass-transfer rate $\dot{M} \sim M_{\rm c}/t_{\dot{a}, \rm CBD} \gg M_{\rm c}/t_{\rm KH}$ could be much greater, favoring instability, e.g., via $L_2$ mass-loss (e.g., \citealt{Lu+23}).  We leave a detailed study of the CBD impact on the binary stability to future work.      

\subsection{Environments of Post-CE Transients}
\label{sec:transients}

\begin{figure}
    \centering
    \includegraphics[width=0.48\textwidth]{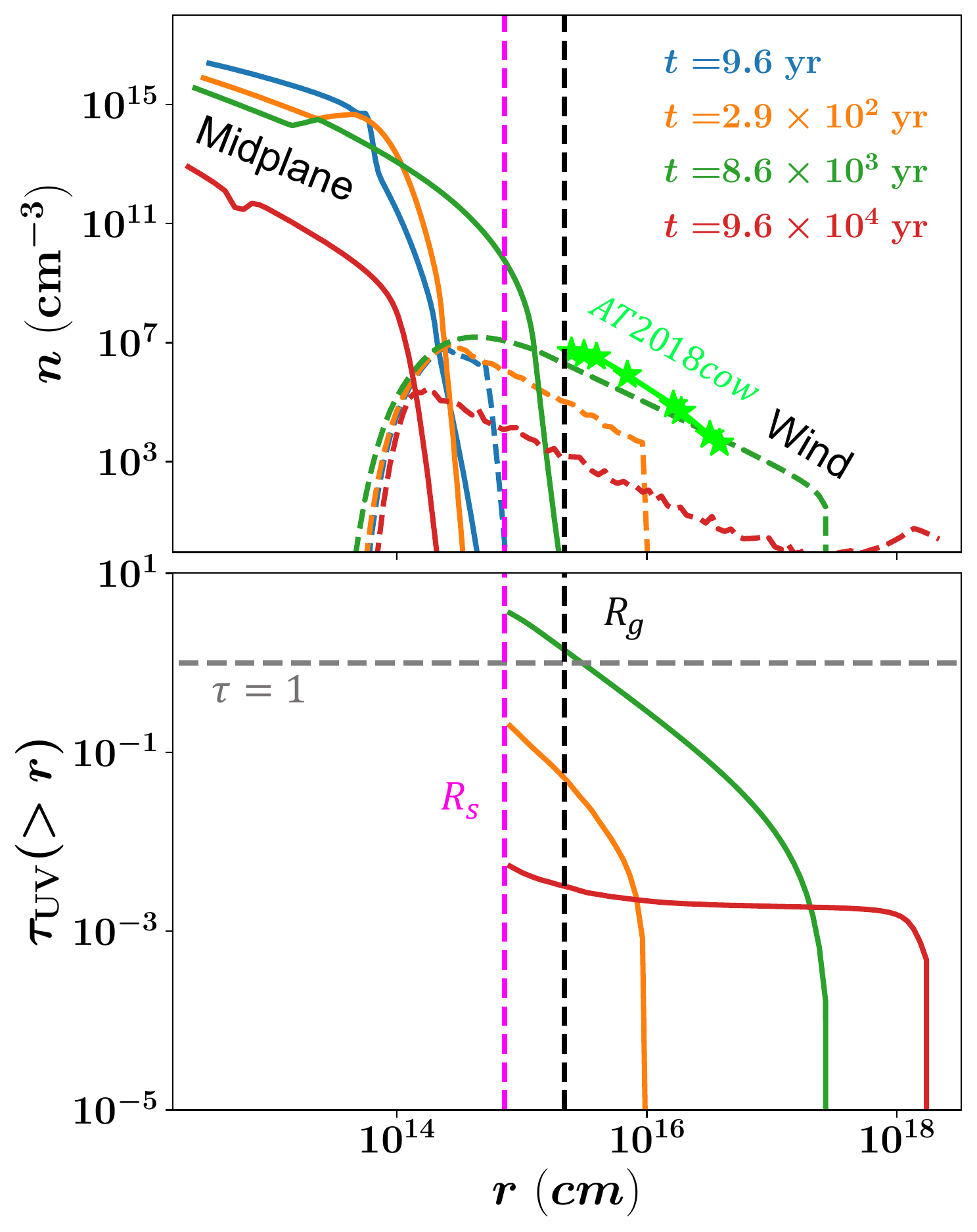}
    \caption{\emph{Top panel:} Snapshots of the radial distribution of particle number density of bound disk material (solid curves) and the outflowing photoevaporation wind (dashed curves, see Eq.~\eqref{eq:rhowind}) for our fiducial model.  For the disk material, we show a spherically-averaged density defined according to $\overline{\rho} \equiv \Sigma\,\pi{}r^2/(4\pi{r}^3/3)$, where $n \equiv \overline{\rho}/m_p$ and $m_p$ is the proton mass. The density profile obtained by modeling the radio/mm emission of AT2018cow \citep{Margutti+19}  shown with green stars.  \emph{Bottom panel:} Snapshots of the optical depth (Eq.\eqref{eq:tauUV}) to UV/optical wavelengths external to a given radius $r$ due to scattering off of spherical dust grains of size $a \approx 1\,\mu{m}$ and assumed opacity $\kappa_{\rm UV} = 200\,\rm cm^2\,g^{-1}$.  Shown for comparison with vertical dashed lines are the photoevaporation radius $R_g$ (Eq.\eqref{eq:Rg}; black) and the dust sublimation radius (Eq.~\eqref{eq:Rs}; magenta). Dust grains are formed roughly exterior to $R_s$, so the optical depth profiles begin at $r = R_s$.}
    \label{fig:environment}
\end{figure}

If the inner binary becomes dynamically unstable, the end result is likely to be the tidal disruption of the He core by the BH/NS \citep{Fryer&Woosley98}.  Subsequent accretion of the shredded core material by the compact object, which occurs on the viscous timescale of only hours to days at highly super-Eddington rates \citep{Metzger22}, was hypothesized to give rise to powerful disk outflows and a merger-driven explosive transient \citep{Chevalier12,Soker+19,Schroder+20,Grichener&Soker21,Metzger22}.  Such events may appear as hydrogen-rich superluminous supernovae (e.g., \citealt{Inserra19}) or other luminous optical and high-energy transients, including FBOTs (e.g., \citealt{Drout+14,Arcavi+16,Prentice+18}).   

A hallmark feature of the most luminous engine-powered FBOT class, as evidenced by their optical (e.g., \citealt{Fox&Smith19}) and X-ray (e.g., \citealt{Margutti+19}) spectra, luminous radio/millimeter synchrotron emission (e.g., \citealt{Ho+19,Margutti+19,Coppejans+20,Bright+21}) and infrared dust emission \citep{Metzger&Perley23}, is the presence of extremely dense circumstellar material (CSM) surrounding the progenitor system on a range of radial scales $\gtrsim 10^{14}-10^{16}$ cm.  Broadly similar dense gaseous environments are inferred to surround Type Ibn \citep{Foley+07,Immler+08,Pastorello+15,Hosseinzadeh+17,Pellegrino+22,Maeda&Moriya22} and Type Icn (e.g., \citealt{Gal-Yam+22,Perley+22}), the light curves and spectra of which point to the presence of up to a few solar masses of hydrogen-depleted CSM on radial scales $\sim 10^{14}-10^{15}$ cm surrounding the progenitor star at the time of explosion (e.g., \citealt{Dessart+21}).

For binary mergers that occur promptly after the dynamical inspiral phase, the requisite gaseous environment could in principle be supplied by the prompt CE ejecta itself (e.g., \citealt{Chevalier12,Soker+19,Schroder+20}).  However, even if the merger is delayed for thousands of years or longer (after which the CE ejecta has expanded to much larger radii $\gtrsim 10^{17}$ cm; e.g., \citealt{Dong+21}), the relic CBD and its photoevaporation-driven outflow studied here may still provide a dense CSM environment \citep{Metzger22}.\footnote{Additional H-depleted CSM can be supplied by mass-loss that occurs from the binary $L_2$ point for many orbital periods during early stages of unstable mass-transfer prior to the dynamical plunge (e.g., \citealt{Pejcha+17})}  Such a delayed merger scenario is arguably favored in AT2018cow due to the low mass $\lesssim M_{\odot}$ (e.g., \citealt{Prentice+18,Perley+19}) and hydrogen-depleted composition of the CSM (e.g., \citealt{Fox&Smith19}) as implied by the optical light curve and spectra, respectively.

If unstable mass-transfer leading to merger is avoided, an energetic transient can still occur, once the stripped core undergoes iron core-collapse at the end of its nuclear lifetime (e.g., \citealt{Tauris+15}).  A core spun up through binary tidal interaction (e.g., \citealt{Cantiello+07,Detmers+08,Fuller&Lu22}) and gaseous accretion from the CBD is more likely to retain large angular momentum at the time of collapse, giving birth to a rapidly spinning compact object, such as an hyper-accreting black hole (e.g., \citealt{MacFadyen&Woosley99,Coughlin19}) or millisecond magnetar (e.g., \citealt{Metzger+11}).  As in long-duration gamma-ray bursts, the resulting explosion is likely to be energetic and jetted (e.g., \citealt{Mosta+15,Gottlieb+22}), compared to the relatively spherical or failed explosions from slower rotating massive stars (see also \citealt{Rueda&Ruffini12}).  Depending on the timing of the collapse, our results show that such core collapse explosions will be shrouded by a similarly dense disk/wind gaseous environment as in the merger case, resulting in luminous CSM shock interaction.  Indeed, compact object birth provides a promising alternative to mergers as a model for luminous FBOTs (e.g., \citealt{Quataert+19}), and the CBD provides a natural source of CSM in this scenario as well.  Even non-jetted core-collapse supernovae embedded in a CBD could be classified as Type Ibn/Icn due to CSM interaction (e.g., \citealt{Metzger22}). 

To quantify the time-dependent explosion environment, the top panel of Fig.~\ref{fig:environment} shows for the fiducial model the radial profiles of the density surrounding the binary as a function of time, including both that of the midplane disk material (solid lines) as well as the photoevaporation-driven wind (dashed lines) on larger scales.  We compute the wind mass flux $\dot{M}_{\rm wind}(t,r)$ according to
\begin{eqnarray}
    &&\dot{M}_{\rm wind}(t,r) = \int_0^t\,dt^\prime\,\times\nonumber\\
    && \int_{R_{\rm in}(t^\prime)}^r\,dr^\prime\,2\pi{r}^\prime\,\dot{\Sigma}_{\rm pe}(t^\prime, r^\prime)\delta\left(\frac{r - r^\prime}{c_s} - (t - t^\prime)\right)   
\end{eqnarray}
and the corresponding wind density
\begin{align}
    \rho_{\rm wind}= \frac{\dot{M}_{\rm wind}(t, r)}{4\pi{r}^2 v_{\rm w}},
    \label{eq:rhowind}
\end{align}
where $v_{\rm w} = c_{\rm s,0} \approx 10$ km s$^{-1}$ is the assumed outflow velocity (Sec.~\ref{sec:photoionization}).  The disk evolves roughly monotonically to lower densities as the disk accretes and viscously spreads outwards, its total mass typically peaking at radii $\sim 10^{14}-10^{15}$ cm (e.g., see $R_{\rm 1/2,m}$ in Fig.~\ref{fig:fiducial_crit_radii}).  By contrast, the wind density profile on radial scales $\gtrsim 10^{15}$ cm peaks at intermediate times $t \sim 10^{4}$ yr, when the outer edge of the disk spreads to radii $R_{\rm pe} \sim 0.1R_{\rm g}$ (Fig.~\ref{fig:fiducial_crit_radii}) and the mass-loss rate peaks (Fig.~\ref{fig:fiducial_disk_mass_j}).

The CBD of mass $\sim 0.1-1M_{\odot}$ thus provides a source of CSM interaction on small radial scales $\lesssim 10^{15}$ cm, contributing to luminous shock-powered (X-ray reprocessed) optical emission (e.g., \citealt{Metzger22}) during the first days to weeks of the merger or core-collapse driven explosion which shape the light curves of luminous FBOTs (e.g., \citealt{Fox&Smith19}) and Type Ibn/Icn (e.g., \citealt{Foley+07,Immler+08,Pastorello+15,Hosseinzadeh+17,Pellegrino+22,Maeda&Moriya22}).  While the He-rich ejecta of Type Ibn favor the merger-disruption or explosion of relatively low-mass He cores, the comparatively He-free ejecta of Type Icn \citep{Gal-Yam+22,Perley+22} favor a more massive and/or chemically homogeneous core at the time of demise \citep{Metzger22}.  

The disk-wind provides a source of CSM interaction on larger scales $\gtrsim 10^{15}-10^{16}$ cm, similar to that needed to explain the luminous radio/mm emission in FBOTs.  For comparison in Fig.~\ref{fig:environment} we show the inferred radial density profile surrounding AT2018cow based on modeling its radio/mm synchrotron emission (e.g., \citealt{Ho+19,Margutti+19,Bright+21}); we find reasonable agreement with the predictions of our fiducial model if the merger/explosion were to occur on a timescale $\sim 10^{4}$ yr after the CE for our fiducial model.  The similarity of this timescale to the thermal time of the core and the disk-driven binary migration $t_{\dot{a},\rm CBD}$ (Fig.~\ref{fig:fiducial_timescales}), over which unstable mass-transfer is expected to commence, lends credence to the delayed merger-driven scenario for LFBOTs \citep{Metzger22}.  A core-collapse explosion of the He core is possible on the same timescale of $\sim 10^{3}-10^{5}$ yr after the CE phase in the subset of interacting binaries for which the original giant mass-transfer commenced only after core He depletion (e.g., \citealt{Klencki+20}; their Fig.~3). 

\subsection{Infrared-Bright Massive Post-CE Binaries}
\label{sec:IR}

As in massive protostars (e.g., \citealt{Patel+05}), the photoevaporation-driven outflow of gas from the CBD can be sufficiently dense to form dust outside the sublimation radius.  Indeed, the weeks-long near-infrared emission from AT2018cow \citep{Perley+19} is naturally interpreted as an ``dust echo'' caused by the absorption and re-emission of early UV radiation by the the same dusty CSM later probed by millimeter/radio emission \citep{Metzger&Perley23}.  

Given the density profile $\rho_{\rm wind}(r,t)$ (Eq.~\eqref{eq:rhowind}), the optical depth through the dusty wind of the inner binary to optical/UV radiation external to radius $r$ can be written:
\be
\tau_{\rm UV}(r,t) = \int_{r}^{\infty}\rho_{\rm wind}\kappa_{\rm UV}dr,
\label{eq:tauUV}
\ee
where $\kappa_{\rm UV} \sim 200 a_{\mu}^{-1}$ cm$^{2}$ g$^{-1}$ is the dust opacity for an assumed grain size $a = 1a_{\mu m}\mu m$.  Large dust grains $a \sim 1\mu m$ are motivated, based on the observed dust properties in post-AGB disks (e.g., \citealt{Gielen+07,Gielen+11,Arneson+17}) and estimates of the grain growth time in the wind \citep{Metzger&Perley23}.  Eq.~\eqref{eq:tauUV} is only valid outside the dust sublimation radius for silicate grains (e.g., \citealt{Metzger&Perley23})
\be
R_{\rm s} \approx 4.4\times 10^{14}\,{\rm cm}\,a_{\mu m}^{-1/2}\left(\frac{L_{\rm bin}}{10^{39}\,{\rm erg\,s^{-1}}}\right)^{1/2}\,\label{eq:Rs},
\ee
external to which dust can form.  As shown in the bottom panel of Fig.~\ref{fig:environment}, we find $\tau_{\rm UV}(R_{\rm s}) > 1$ is satisfied for the fiducial model at times $t \sim 10^{4}-10^{5}$ yr.  During this temporal window, a significant fraction of the binary luminosity will be absorbed by the wind near $R_{\rm s}$ and reprocessed into thermal near-infrared (nIR) emission peaking near the sublimation temperature $T \approx 2000$ K at a wavelength of 1-2$\mu m$.

We now estimate the detectability of optically-shrouded but nIR-bright post-CE binary systems with the {\it Nancy Grace Roman Space Telescope} (hereafter {\it Roman}; \citealt{Spergel+15}).  The planned High Latitude Wide Area Survey (HL-WAS) using the Wide Field Instrument (WFI) is expected to cover $\Omega = $2000 deg$^{2}$ (a fraction $f_{\Omega} = 0.048$ of the full sky) in the wavelength range $\lambda \approx 1-2\mu$m down to an AB magnitude depth of 26.5 mag ($F_{\rm lim} \gtrsim 10^{-30}$ erg cm$^{-2}$ s$^{-1}$).  The detection horizon corresponding to a $1\mu m$ luminosity of $L_{\rm bin} \approx \nu L_{\nu}$ is thus given by 
\be
D_{\rm lim} \simeq \left(\frac{L_{\nu}}{4\pi F_{\rm lim}}\right)^{1/2} \approx 0.16\,{\rm Gpc}\,\left(\frac{L_{\rm bin}}{10^{39}\,{\rm erg\,s^{-1}}}\right)^{1/2}.
\label{eq:Dlim}
\ee
Given the estimated volumetric rate of BH/NS common envelope events in the local universe $\mathcal{R} \sim 10^{2}-10^{3}$ Gpc$^{-3}$ yr$^{-1}$ (e.g., \citealt{Vigna-Gomez+18,Schroder+20,Grichener23}), we therefore estimate the total number of sources in the HL-WAS:
\begin{align}
&N_{\rm HL-WAS} \simeq \frac{4\pi}{3}f_{\Omega}D_{\rm lim}^{3}\mathcal{R}t_{\tau > 1}\nonumber\\
&\approx 
10^{3} \left(\frac{L_{\rm bin}}{10^{39}\,{\rm erg\,s^{-1}}}\right)^{3/2}\left(\frac{\mathcal{R}}{100\,\rm Gpc^{-3}yr^{-1}}\right)\left(\frac{t_{\tau > 1}}{10^{4}\,{\rm yr}}\right)
\label{eq:NHLWAS}
\end{align}
where $t_{\tau > 1} \sim 10^{4}$ yr is the estimated lifetime of the nIR source (e.g., the time interval over which $\tau > 1$ in Fig.~\ref{fig:environment}).  The number of detectable systems would be higher if we considered all post-CE massive binaries, i.e. including also those with main-sequence companions.

Depending on the luminosity function and rates of shrouded events, we conclude that a survey similar to HL-WAS with {\it Roman} will detect up to thousands of post-CE binaries, i.e., candidate Type Ibn/Icn or FBOT progenitors.  At distances of $\lesssim $ 100 Mpc, the characteristic angular size of $L_{\star}$ galaxies is a few arcseconds (e.g., \citealt{Metzger+13}), compared to the $0.11$ arcsec per pixel resolution of the WFI, thus such luminous individual point sources to be identified at this distance.  However, optical/NIR spectra and follow-up observations at other wavelengths (particularly radio; see Sec.~\ref{sec:hypernebulae}), may be necessary to distinguish these objects from other IR-bright sources, such as as massive protostars (e.g., \citealt{Jones+22}) or other dust-enshrouded supernovae progenitors of a distinct origin (e.g., SN 2008S or the NGC 300 transient; \citealt{Thompson+09}).

\subsection{Post-CE Radio Hypernebulae}
\label{sec:hypernebulae}

Accretion onto the BH/NS from the CBD may power a long-lived jet in the post-CE phase (e.g., \citealt{Balick&Frank02,Soker02}), somewhat akin to the binary systems which power ultraluminous X-ray source (ULX; \citealt{Kaaret+17}) during earlier pre-CE mass-transfer phases (e.g., \citealt{Pavlovskii+17}). This picture is supported in lower-mass post-CE binaries by direct observational evidence for bipolar jets or collimated fast disk outflows in disk-bearing post-AGB systems (e.g., \citealt{Thomas+13,Gorlova+15}).

The accretion rate onto the inner binary, and hence onto the mini-disk feeding the BH/NS, is generally highly super-Eddington (Fig.~\ref{fig:fiducial_disk_mass_j}).  The inner portions of the disk approaching the compact object are therefore susceptible to powerful outflows which carry away a substantial fraction of the inflowing mass (e.g., \citealt{Blandford&Begelman99}).  We account for this reduction in the mass-accretion rate which reaches innermost stable circular orbit (ISCO) of the BH (or surface radius of the NS, which is similar in size) according to
\be
\dot{M}_{\bullet} = \dot{M}(R_{\rm in})\left(\frac{R_{\rm isco}}{R_{\rm trap}}\right)^{p},
\label{eq:MdotBH}
\ee
where $R_{\rm isco} = 6R_{\rm G}$ is the ISCO radius for a slowly spinning BH and a value of $p = 0.6$ is motivated by numerical simulations of radiatively inefficient accretion flows (e.g., \citealt{Yuan&Narayan14}). 

The bottom panel of Fig.~\ref{fig:fiducial_disk_mass_j} shows the BH accretion rate $\dot{M}_{\bullet}$ evolution for the fiducial model.  Although the trapping radius moves inside the binary orbit (equivalently, $\dot{M}_{\rm in} < \dot{M}_{\rm edd,in}$) within the first few years of disk evolution, we see that $\dot{M}_{\bullet} \gtrsim \dot{M}_{\rm edd}$ for almost the entire evolution of the disk.  During this sustained phase of super-Eddington accretion, a bipolar relativistic jet from the BH/NS (e.g., \citealt{Sadowski&Narayan15}) may be produced which carries a luminosity
\be
L_{\rm j} = \eta_{\rm j} \dot{M}_{\bullet}c^{2},
\ee
where $\eta_{\rm j} \sim 0.1-1$ is an efficiency factor that depends on details such as the magnetic flux of the accreted material (e.g., \citealt{Tchekhovskoy+10}).  
Over the duration of the post-CE super-Eddington accretion phase, we estimate that such a jet will release an energy $E_{\rm j} = \int L_{\rm j}dt \sim 2\times{10}^{52}\,\eta_{\rm j} {\rm erg}$ over a timescale $\sim 10^{4}-10^{5}$ yr.  

The transient jet will inflate a several-parsec-sized bubble of plasma and magnetic fields, akin to the ``hypernebulae'' envisioned to accompany Roche-lobe overflow preceding a CE (e.g., \citealt{Sridhar&Metzger22}).  Following \citet{Sridhar&Metzger22} we estimate that for a jet with $\dot{M}/\dot{M}_{\rm Edd} \sim 10$ and duration $\sim 10^{4}$ yr, the jet/wind-driven bubble will expand to a maximum size of a few pc and achieve a 1 GHz radio luminosity $\nu L_{\nu} \sim 0.01-0.1L_{\rm j} \sim 10^{37}-10^{38}$ erg s$^{-1}$, comparable to bright radio supernovae.  This corresponds to a readily detectable flux density $F_{\nu} \gtrsim 1-10$ mJy for a characteristic source distance $\lesssim 0.1$ Gpc (Eq.~\eqref{eq:Dlim}) of putative nIR-discovered candidate systems (Sec.~\ref{sec:IR}).  

On the other hand, the total mass accreted by the compact object $\lesssim 0.1 M_{\odot}$ during the CBD phase is unlikely to significantly grow the mass of the compact object.  We do not therefore expect this phase to create unexpectedly massive NS or low-mass BH, such as those required to explain the LIGO binary mergers GW190425 \citep{Abbott+20} or GW190814 \citep{Abbott+20b}, by accretion onto a canonical (e.g., $1.4M_{\odot}$) mass NS.

\section{Conclusions}
\label{sec:conclusions}

We have developed a parameterized one-dimensional height-integrated model for the long-term evolution of a circumbinary accretion disk created from a (partially-)failed massive-star CE event and explored the resulting implications for the final state of the binary and the role of its time-dependent gaseous environment on the appearance of post-CE systems and associated explosive transients.  Our results can be summarized as follows:

\begin{enumerate}[label=(\roman*)]

\item As the CBD evolves under the effects of viscous angular momentum transport, it undergoes a wide-range of physical conditions spanning those of ultraluminous X-ray binaries at small radii to those of massive proto-star disks at large radii (Figs.~\ref{fig:paramspace}, \ref{fig:fiducial_snapshots}).  At early times and small radii, the disk is hot, radiation-dominated, geometrically- and optically-thick, and heated primarily by viscous dissipation.  As the disk spreads, it becomes thinner, gas-pressure dominated, and heated primarily by irradiation from the luminosity of the central binary (Fig.~\ref{fig:fiducial_disk_fractions_snapshots}).  The details of the evolution and its overall timescale depend on the assumed parameters (particularly, the viscosity $\alpha$ and the initial disk mass and angular momentum), but the overall evolution is robust to these uncertainties (Fig.~\ref{fig:non_fiducial}).   

\item The ultimate dispersal of the disk results from a combination of accretion onto the central binary and photoevaporation mass-loss from large radii (Figs.~\ref{fig:fiducial_disk_mass_j}, \ref{fig:non_fiducial}).  The characteristic disk lifetime $\sim 10^{4}-10^{5}$ yr can be comparable to the thermal timescale of the He core, and even to the nuclear timescale for a highly-evolved core.  These lifetimes are also similar to those of disks around massive proto-stars (e.g., \citealt{Fuente+06,Beltran&deWit16}), perhaps not coincidentally given the similar physics involved (e.g., \citealt{Hollenbach+00}).

\item The magnitude (and even sign) of the binary torques due to accretion from the CBD remain poorly understood theoretically, particularly across the wide and time-dependent range of conditions characterizing the inner boundary of the CBD.  However, for values of $\chi$ spanning those found by recent CBD simulation suites (e.g., \citealt{Zrake+21,Siwek+23}), significant evolution of the binary separation can occur for disk masses which correspond to even a modest residual envelope mass being left from the CE phase (e.g., $f_{\rm m} \gtrsim 0.1$; Fig.~\ref{fig:delta_alpha_CE}), as follows analytically from Eq.~\eqref{eq:adotbin2}.  The corresponding change to the {\it effective} value of $\alpha_{\rm CE}$ relative to CBD-less case ranges from tens of percent to a factor of two or more (Fig.~\ref{fig:delta_alpha_CE}).  Values of $\alpha_{\rm CE} > 1$ which are forbidden within the usual CE formalism in principle become possible when a CBD is present.

\item The possibility that the CBD lifetime can exceed the thermal (or even nuclear) timescale of the core implies that the disk may well be present during any second mass-transfer phase between the core and compact object companion.  If the mass-transfer process commences prior to the core thermal times, binary torques from the CBD could significantly enhance the core's mass-loss rate through $L_1$ (or $L_2$) relative to an isolated binary transferring mass on the thermal timescale (Fig.~\ref{fig:fiducial_timescales}).  Even if the CBD exerts a negligible torque on the binary after mass-transfer begins, the inflowing mass streams may affect the binary evolution indirectly by disrupting how the BH/NS mini-disk feed angular momentum back into the orbit.  Binary eccentricity excited by the CBD could also affect the mass-transfer stability. 

\item Unstable mass-transfer between the He core and the BH/NS can lead to tidal disruption and an explosive transient (e.g., \citealt{Soker&Gilkis18,Schroder+20,Metzger22}), which$-$due to the relatively gradual evolution of the CBD$-$can be delayed by thousands of years or longer after the CE event.  Such delayed mergers will not take place in a vacuum: the CBD (on small scales $\lesssim 10^{14}$ cm) and its photo-evaporation driven wind (on large scales $\gtrsim 10^{15}$ cm) provide a dense gaseous medium into which the fast ejecta from the merger will collide (Fig.~\ref{fig:environment}), generating luminous electromagnetic radiation.  The radial density profiles implied by our models are broadly consistent with those observed to surround luminous FBOTs such as AT2018cow \citep{Ho+19,Margutti+19,Fox&Smith19,Metzger&Perley23} and Type Ibn/Icn supernovae \citep{Foley+07,Dessart+21,Gal-Yam+22,Perley+22}.

\item The same dense CSM surrounding the CBD is likely to generate dust which is opaque to UV radiation from the central He star and BH/NS disk (Fig.~\ref{fig:environment}).  The binary's luminosity will thus be reprocessed into nIR emission, rendering post-CE systems potentially promising sources for future discovery with the {\it Roman Space Telescope} out to characteristic distances $\sim 100$ Mpc.  We also encourage searches for excess nIR emission from suspected post-CE binaries discovered with UV surveys (e.g., \citealt{vanSon+22}).  

\item During the CBD phase the BH/NS is accreting at or above the Eddington rate (Fig.~\ref{fig:fiducial_disk_mass_j}, bottom panel), qualitatively similar to the ULX phase which can precede massive-star CE events.  As in the case of ULX like SS433, the accreting BH/NS will launch a fast disk wind or relativistic jet of plasma, inflating a multi-parsec scale synchrotron nebula surrounding the binary (e.g., \citealt{Sridhar&Metzger22}).  Such off-nuclear nIR-luminous radio sources could help identify the post-CE CBD systems which contain compact object accretors.  

\end{enumerate}

\acknowledgements

We thank Joe Bright, Jared Goldberg, Zoltan Haiman, Jakub Klencki, Raffaella Margutti, Ondrej Pejcha, and Mathieu Renzo, Lieke van Son for useful conversations and helpful information.  S.T. and B.D.M. are supported in part by the NSF through the NSF-BSF grant program (grant AST-2009255).  The Flatiron Institute is supported by the Simons Foundation.

\appendix

\section{Analytic Estimates of Critical Radii}
\label{app:analytic}

Across a wide radial range, the disk is gas-pressure-supported, irradiation-heated and cools radiatively.  The radial profiles of the midplane temperature and aspect ratio of the disk under these conditions are obtained by balancing the rates of heating and radiative cooling in Eq.~\eqref{eq:thermal_balance}.  This gives:
\begin{align}
    T &\approx 5\,\times\,10^3\,K\,\left(\frac{\mu}{2}\right)^{-1/7}\left(\frac{L_{\rm bin}/M_{\rm bin}}{1.38\,\times\,10^{38}\,\text{erg}\,\text{s}^{-1}\,M_\odot^{-1}}\right)^{2/7}\left(\frac{M_{\rm bin}}{20  M_\odot}\right)^{1/7}\left(\frac{r}{200 R_\odot}\right)^{-3/7}\\
    \theta &\approx 0.03\,\left(\frac{\mu}{2}\right)^{-4/7}\left(\frac{L_{\rm bin}/M_{\rm bin}}{1.38\,\times\,10^{38}\,\text{erg}\,\text{s}^{-1}\,M_\odot^{-1}}\right)^{1/7}\left(\frac{M_{\rm bin}}{20  M_\odot}\right)^{-3/7}\left(\frac{r}{200 R_\odot}\right)^{2/7}
\end{align}
Equating the rates of irradiation heating and viscous heating,
\begin{align}
    \frac{L_{\rm bin}}{2\pi\,r^2}\theta = \frac{9}{4}\nu\Omega^2\Sigma\,f(\tau),
\end{align}
we obtain an estimate of the critical radius external to which irradiation heating dominates:
\begin{align}
    R_{\rm irrad, 1/2} \approx 700\,R_\odot\left(\frac{M}{20\,M_\odot}\right)^{1/9}\left(\frac{L_{\rm bin}/M_{\rm bin}}{1.38\times{10}^{38}\,M_\odot^{-1}}\right)^{-2/9}\left(\frac{\alpha}{0.01}\right)^{2/9}\left(\frac{\kappa}{0.001\,\rm cm^2\,g^{-1}}\right)^{2/9}\left(\frac{\theta}{0.1}\right)^{2/9}\left(\frac{\Sigma_0}{10^6\,\rm g\,cm^{-2}}\right)^{4/9},
\end{align}
where we have taken $f(\tau) \approx (3/2)\tau$ appropriate to the optically-thick limit.

At small radii, viscous heating balances radiative cooling and the disk is optically thick $\tau \gg 1$.  Under these conditions, we find that radiation pressure exceeds gas pressure interior to the critical radius
\begin{align}
    R_{\rm rad, 1/2} \approx 500\,R_\odot\left(\frac{M}{20\,M_\odot}\right)^{1/35}\left(\frac{R_{\rm d,0}}{200\,R_\odot}\right)^{32/35}\left(\frac{\alpha}{0.01}\right)^{2/5}\left(\frac{\kappa}{1\,\rm cm^2\,g^{-1}}\right)^{2/5}\left(\frac{\mu}{2}\right)^{-8/35}\left(\frac{\Sigma_0}{10^6\,\rm g\,cm^{-2}}\right)^{16/35}.
\end{align}
This expression is consistent with the purple contour $R_{\rm rad, 1/2} \propto \Sigma_0^{0.5}$ in Fig.~\ref{fig:paramspace} at radii $30\,R_\odot \lesssim r \lesssim 10^3\,R_\odot$, given that the opacity is relatively constant across this region.

When radiation pressure dominates, advective cooling and radiative cooling become equal at a critical vertical aspect ratio $\theta \approx \sqrt{3}/2$.  Substituting this into $F_{\rm visc} \approx P_{\rm rad}c/(2\kappa\Sigma)$, we find that advective cooling dominates interior to a critical radius
\begin{align}
    R_{\rm adv, 1/2} \approx 480\,R_\odot\left(\frac{M}{20\,M_\odot}\right)^{1/5}\left(\frac{R_{\rm d,0}}{200\,R_\odot}\right)^{4/5}\left(\frac{\alpha}{0.01}\right)^{2/5}\left(\frac{\kappa}{1\,\rm cm^2\,g^{-1}}\right)^{2/5}\left(\frac{\Sigma_0}{10^6\,\rm g\,cm^{-2}}\right)^{2/5}.
\end{align}
Although the normalization of this estimate slightly overestimates the olive contour in Fig.~\ref{fig:paramspace}, the scaling with $\Sigma_0$ is roughly consistent.

The disk can also become radiation pressure dominated at large radii, where irradiation dominates the heating rate. At these radii, we have 
\begin{align}
 \frac{L_{\rm bin}}{4\pi r^2}\theta \approx \sigma{T}^4
\end{align}
Equating $P_{\rm gas} = P_{\rm rad}$ in the region $\tau > 1$, we obtain
\begin{align}
    \Sigma \approx \frac{4}{3GM\pi\Sigma}
\end{align}
which corresponds to
\begin{align}
    R_{\rm rad, 1/2} \approx 5.2\times\,10^4\,R_\odot\left(\frac{L_{\rm bin}/M_{\rm bin}}{1.38\times{10}^{38}\,M_\odot^{-1}}\right)^{-1/2}\left(\frac{\Sigma_0}{10^6\,\rm g\,cm^{-2}}\right)^{1/2}\left(\frac{R_{\rm d,0}}{200\,R_\odot}\right)
\end{align}

\section{Summary of Disk-Binary Evolution Equations and Solution Method}
\label{app:summary_equations}

For convenience, we collect here the equations that we solve for the coupled disk/binary evolution:
\begin{align}
    \frac{\partial\Sigma}{\partial{t}} &= \frac{3}{r}\frac{\partial}{\partial{r}}\left(r^{1/2}\frac{\partial}{\partial{r}}(\nu\Sigma{r^{1/2}})\right) - \dot{\Sigma}_{\rm pe}\label{eq:sigma_ev2}\\
    \frac{\dot{a}_{\rm bin}}{a_{\rm bin}} &=  \left[2\chi\frac{(q+1)^{2}}{q} - 3 \right]\frac{\dot{M}_{\rm bin}}{M_{\rm bin}}\\
    \dot{M}_{\rm bin} &= -\dot{M}_{\rm in} = -2\pi\,R_{\rm in}(t)\Sigma\,v_{\rm in}\label{eq:mdot_bin}\\
    R_{\rm in}(t) &= 2a_{\rm bin}(t)\\
    v_{\rm in} &= -\frac{R_{\rm in}}{a_{\rm bin}}\dot{a}_{\rm bin} - \frac{3}{\Sigma r^2\Omega}\frac{\partial}{\partial r}\left(\nu\Sigma{r^2\Omega}\right)\bigg\rvert_{r = \rm R_{\rm in}(t)}\label{eq:bdry2}
\end{align}
where $\dot{\Sigma}_{\rm pe}$ is given by Eq.\eqref{eq:sigma_dot_pe}-\eqref{eq:rho_pe}, and $\nu(\Sigma, r) = \alpha(P/\rho \Omega)$ is obtained from the vertical structure equations:
\begin{align}
    \Sigma &= 2\rho H\\
    \theta &= \frac{h}{r}\\
    \kappa &= \kappa(\rho, T)\\
    \tau &= \frac{\kappa\Sigma}{2}\\
    P_{\rm gas} &= \frac{k\rho T}{\mu m_p}\\
    P_{\rm rad} &= (1 - e^{-\tau^2})\frac{a T^4}{3}\\
    \Omega &= \left(\frac{GM_{\rm bin}}{r^3}\right)^{1/2}\\
    P_{\rm gas} + P_{\rm rad} &= \rho h^2 \Omega^2\\
    \frac{9}{4}\nu\Omega^2\Sigma  + \frac{L_{\rm bin}}{2\pi\,r^2f(\tau)}\theta &= \frac{2\sigma T^4}{f(\tau)} + \xi\frac{3\nu}{2r}\theta\left(\frac{3}{2}P_{\rm gas} + 4P_{\rm rad}\right)\\
    f(\tau) &= \frac{3}{2}\tau + 2 + \frac{1}{(\tau + \tau_{\rm min})}
\end{align}
with $\tau_{\rm min} = 0.01$, $\mu = 2$ and the opacity $\kappa(\rho, T)$ is from \citet{Bell&Lin94} as implemented in the radiation model of the PLUTO code \citep{PlutoCode}.

We first solve the vertical structure equations for the midplane temperature $T$ and aspect ratio $\theta$ on a logarithmically spaced grid on the $\Sigma - r$ plane, and numerically construct $\nu(\Sigma, r)$.  We then use this mapping to evolve Eqs.~\eqref{eq:sigma_ev2}-\eqref{eq:bdry2} forward in time, starting from the given initial binary properties and disk surface density profile. Our tabulated function $\nu(\Sigma, r)$ assumes $M_{\rm bin} = M_{\rm bin,0}$ (although we evolve $M_{\rm bin}(t)$ through Eq.~\eqref{eq:mdot_bin}), which introduces a negligible error in the solution, since $M_{\rm bin}$ changes by at most a few percent during its evolution, since $M_{d,0} \ll M_{\rm bin}$ for the range of disk masses we considered.

We found it convenient to use a static radial grid in the numerical calculation; hence, we first apply a time-dependent shift in the radial coordinate,
\begin{align}
    \tilde{r}(r, t) &\equiv r + 2\left(a_{\rm bin}(0) - a_{\rm bin}(t)\right),   
\end{align}
so that $\tilde{r} \in[2a_{\rm bin}(0), \infty)$, thus mapping the inner boundary of the disk to its initial radius at all times. We further change coordinates to dimensionless variables
\begin{align}
    \overline{r}(\tilde{r}) &\equiv \left(\frac{\tilde{r}}{R_{d,0}}\right)^m\\
    \tilde{t}(t) &\equiv \left(\frac{t}{t_{\nu, 0}}\right)^n
\end{align}
where $R_{d,0}$ is the characteristic length scale for the initial data defined in Eq.~\eqref{eq:init_data}, $t_{\nu,0}$ is the viscous timescale at this radius given by $t_{\nu,0} = (2/3)R_{d,0}^2/\nu(R_{d,0}, \Sigma(t= 0, R_{d,0}))$. We discretize the system of equations Eq.~\eqref{eq:sigma_ev2} - ~\eqref{eq:bdry2} on a uniform grid in the coordinate $\overline{r}$. We (arbitrarily) set $m= n = 1/3$ in most of our simulations to increase resolution at early times and inner radii where the evolution of the disk is most dynamical. The diffusion equation we actually solve is, therefore,
\begin{align}
    \frac{\partial\Sigma}{\partial{\tilde{t}}} = \frac{dt}{d\tilde{t}}\left(\frac{3}{r}\frac{d\overline{r}}{d\tilde{r}}\frac{\partial}{\partial\overline{r}}\left(r^{1/2}\frac{d\overline{r}}{d{\tilde{r}}}\frac{\partial}{\partial\overline{r}}\left(r^{1/2}{\nu}{\Sigma}\right)\right) -\dot{\Sigma}_{\rm pe} + 2\dot{a}_{\rm bin}\frac{d\overline{r}}{d{\tilde{r}}}\frac{\partial\Sigma}{\partial\overline{r}}\right), 
\end{align}
where all functions are evaluated at $(t,r) = (t(\tilde{t}), r(\tilde{r}(\overline{r}, \tilde{t}))$. We place the outer boundary at $R_{\rm out} = 10^6\,R_\odot$. For the fiducial model (Table~\ref{tab:parameters_symbols}), $R_{\rm in}(0) = 60\,R_\odot$, and $R_{d,0} = 238\,R_\odot$. This is mapped to $\overline{r} \in[0.63, 16]$, which we represent by $512$ equally spaced grid points. We set a maximum time step $\Delta\tilde{t} = 0.01$, and allow our solver to take smaller steps when the root finding algorithm fails for the given value of $\Delta\tilde{t}$. We terminate simulations when disk total disk mass drops to $10^{-5}$ of its initial value.

We use the second-order accurate, and implicit Crank-Nicholson \citep{Press2007} finite-difference method. We use Scipy's \citep{2020SciPy-NMeth} optimization routine ``root" \citep{Mor1980UserGF} to solve the algebraic equations arising from finite difference approximation.

\bibliographystyle{aasjournal}
\bibliography{refs}

\end{document}